\documentclass{aastex63}

\definecolor{TodoColor}{RGB}{239, 0, 101}

\usepackage{multirow}
\usepackage[section]{placeins}
\usepackage[normalem]{ulem}

\shorttitle{The Connection Between GALEX and Kepler Flares}
\shortauthors{Brasseur et al.}

\graphicspath{{./}{figures/}}

\begin{document}

\title{Constraints on Stellar Flare Energy Ratios in the NUV and Optical From a Multiwavelength Study of GALEX and Kepler Flare Stars}

\correspondingauthor{C.E.  Brasseur}
\email{cb423@st-andrews.ac.uk}

\author[0000-0002-9314-960X]{C.E.  Brasseur}
\affiliation{Space Telescope Science Institute, 3700 San Martin Dr, Baltimore, MD 21218, USA}
\affiliation{SUPA, School of Physics and Astronomy, University of St. Andrews, Fife, KY16 9SS, Scotland}

\author[0000-0001-5643-8421]{Rachel A. Osten}
\affiliation{Space Telescope Science Institute, 3700 San Martin Dr, Baltimore, MD 21218, USA}
\affiliation{Center for Astrophysical Sciences, Department of Physics \& Astronomy, Johns Hopkins University, 3400 North Charles St., Baltimore, MD 21218, USA}

\author[0000-0001-5974-4758]{Isaiah I. Tristan}
\affiliation{University of Colorado Boulder, 3665 Discovery Drive, Boulder, CO 80303}

\author[0000-0001-7458-1176]{Adam F. Kowalski}
\affiliation{University of Colorado Boulder \& The National Solar Observatory, 3665 Discovery Drive, Boulder, CO 80303}

\begin{abstract}
We present a multiwavelength study of stellar flares on primarily G-type stars using overlapping time domain surveys in the near ultraviolet (NUV) and optical regimes. The NUV (GALEX) and optical (Kepler) wavelength domains are important for understanding energy fractionations in stellar flares, and for constraining the associated incident radiation on a planetary atmosphere. We follow up on the NUV flare detections presented in \citet{Brasseur_2019ApJ...883...88B}, using coincident Kepler long (1557 flares) and short (2 flares) cadence light curves. We find no evidence of optical flares at these times, and place limits on the flare energy ratio between the two wavebands. We find that the energy ratio is correlated with GALEX band energy, and extends over a range of about three orders of magnitude in the ratio of the upper limit of Kepler band flare energy to NUV flare energy at the same time for each flare. The two flares with Kepler short cadence data indicate that the true Kepler band energy may be much lower than the long cadence based upper limit. A similar trend appears for the bulk flare energy properties of non-simultaneously observed flares on the same stars. We provide updated models to describe the flare spectral energy distribution from the NUV through the optical including continua and emission lines to improve upon blackbody-only models. The spread of observed energy  ratios is much larger than encompassed by these models and suggests new physics is at work. These results call for better understanding of NUV flare physics and provide a cautionary tale about using only optical flare measurements to infer the UV irradiation of close-in planets.

\end{abstract}

\keywords{}

\section{Introduction} \label{sec:intro}

Flares are the most dramatic energy release events that cool stars will experience while on the main sequence. They occur as a result of magnetic reconnection, which involves non-linearities in magnetic field and plasma coupling. Stellar flares are one manifestation of magnetic activity which is demonstrated by all solar-like stars (as well as stars across many dwarf classifications) to varying degrees and can be observed across the majority of the electromagnetic spectrum. Flares have recently become a prime topic of concern for estimating the true habitability of planets around M dwarfs, where the close proximity of the habitable zone yields high surface fluxes of ionizing radiation for orbiting planets \citep{Howard2020}. Flares may have a positive impact on planets surrounding M dwarf stars by increasing the light at UV wavelengths needed for UV-sensitive prebiotic chemistry  \citep{Ranjan2017} as well as augmenting the effectiveness of photosynthesis at visible wavelengths \citep{Mullan2018}. Even for solar analogue stars, the range of observed flare variations is far larger than historical records for the Sun indicate \citep{Maehara2012} and require investigation into effects on nearby planets \citep{Airapetian2020}.  Because of the dramatic increases seen at ultraviolet and optical wavelengths during stellar flares, it is particularly important to understand the range of flare energies in those wavelengths, in order to quantify the effect of flares on close-in exoplanets \citep{Ranjan2017}. The occurrence rate of stellar flares generally follows that of other stellar magnetic activity indicators (e.g. young stars with fast rotation and enhanced X-ray emission also have a high flaring rate).  While our Sun is a star that can be studied in amazing spatial, temporal and spectral resolution, it is only one star observed at a singular point in its evolutionary history. Diversifying flare studies to a wider intrinsic variety of stellar types demonstrate the extent to which the Sun's activity manifestation represents the real variation that can be expected from stellar flares of all types.

Flares are detectable as rapid increases in stellar intensity across the electromagnetic spectrum, from radio wavelengths to high energy gamma rays \citep{Osten_2005ApJ...621..398O,Osten_2016ApJ...832..174O}. This demonstrates the range of physical processes involved in stellar  flares (plasma heating, particle acceleration, shocks and mass flows; \citet{Osten2016}), as well as the involvement of all stellar atmospheric layers in the flare, from the dense photosphere to the high temperature, rarefied corona. Flaring seen at ultraviolet wavelengths originates from the chromosphere/transition region and also includes the response of the photosphere to the input of energy. Other magnetic activity indicators (e.g. H$\alpha$, X-ray radiation) produce emission during quiescence and flares, whereas the enhanced blue optical continuum radiation is not seen outside of flares, making it a key signature of the flare process \citep{Osten2016}.

Detailed studies of solar and stellar flares show the ubiquity of continuum enhancements at ultraviolet through optical wavelengths \citep{Kretzschmar_2011A&A...530A..84K, Kowalski_2013ApJS..207...15K}. This is often characterized phenomenologically as a blackbody with a temperature near 10$^{4}$ K. 
\citet{Kowalski_2013ApJS..207...15K} pointed out the presence of an additional continuum component visible at wavelengths longer than H$\beta$ ($\lambda \ge 4900$ \AA) which increases in importance to the flare energy budget during the late gradual phase. The Kepler bandpass, extending from 4000-9000 \AA \citep{Kepler_paper, Kepler_bandpass_ref}, encompasses most of this continuum emission, and disentangling the amount of hot blackbody radiation from red continuum emission requires constraints at near ultraviolet (NUV) wavelengths as well as modeling. Radiative hydrodynamic models, such as those of \citet{Kowalski_2017ApJ...837..125K}, which constrain the hot blackbody and predict optical flare line and Balmer emission, are key to interpreting the emission processes occurring in the optical. The Galaxy Evolution Explorer (GALEX) NUV bandpass covers the short wavelength side of this presumed blackbody and contribution from the Balmer continuum, while the Kepler (and TESS) bandpass extends to the red, \citep{galexref}. 

Ultraviolet (UV) fluxes are formed in cool stars from line and continuum emission from the photosphere, chromosphere, and transition region, and their formation originates from magnetic heating processes not predictable from stellar photospheric models alone \citep{Benz_Gudel_2010ARA&A..48..241B}. These wavelengths are critical for changing the atmospheric chemistry on planets in the habitable zone. Depending on the nature of the exoplanet atmosphere, as much as 4\% of the incident UV radiation may penetrate the exoplanet atmosphere to regions below \citep{Smith_2004OLEB...34..513S}. In standard flare models the UV spectral region carries the bulk of the radiated energy  \citep{Hawley_1995ApJ...453..464H, Osten_Wolk_2015ApJ...809...79O}. A steady-state chemistry in the planet's atmosphere is driven by the star's quiescent UV radiation field \citep{France_2014Ap&SS.354....3F}, but frequent UV flares may alter it. For example, \citet{Venot_2016ApJ...830...77V} found that recurrent flares affect the atmospheric chemistry of exoplanets, particularly transmission spectra taken at different times relative to the flares; the magnitude of this effect will be observable with JWST. Thus, quantifying the occurrence rate of stellar UV flares with differing flare energies is a vital input to constraining the range of possibilities for exoplanet chemistry. \citet{Tilley_2017arXiv171108484T} recently included the effect of frequent stellar flares on exoplanet atmospheric chemistry, specifically the potential for it to destroy planetary ozone shields. \citet{Segura_2010AsBio..10..751S} and \citet{Ranjan_2017AsBio..17..169R} model the effect of stellar flares on exoplanet atmospheric chemistry, assuming that the NUV flare spectrum is uniform throughout the whole flare and among all flares. Quantifying the range of variability will be key to updated inputs to more realistic modeling. In light of the expected dearth of UV space telescopes sometime in the current decade, with the expected demise of Hubble and Swift, along with the continued emphasis on precision optical timing measurements of stars to detect transiting planets, it is useful to have constraints on the amount of energy emitted by a flare in the NUV and optical bandpasses from simultaneous observations.

The current sparse multi-wavelength data on stellar flares exhibit correlations and energy partitions similar to those seen in solar flares, suggesting that the same physical process is occurring in both \citep{Osten_Wolk_2015ApJ...809...79O}, but detailed studies can reveal the extent of the similarities. This is necessary to understand the connection between the well-studied solar flares (reaching a maximum of about $10^{32}$ erg), and the much more energetic events seen on other stars (flares on nearby single stars with well-constrained parameters can reach to $10^{36}$ erg). Despite the tremendous advantage to studying flares at both wavelengths at the same time, such multi-wavelength observations are difficult to arrange, and few exist. This study takes advantage of the temporal and spatial overlap between the GALEX (NUV) and Kepler (optical) space telescope missions between 2009 and 2013. Because both were survey missions not focussed on flare detection, they give a wealth of information on stars that are not inherently high flaring rate stars, but offer an excellent opportunity for serendipitous flare detections. Several previous studies \citep{Davenport_2016ApJ...829...23D,vandoors,Yang_2019ApJS..241...29Y} have explored flares in Kepler data, and our previous study \citep{Brasseur_2019ApJ...883...88B} describes a body of small short-duration flares in the coincident GALEX data. This paper combines data from the two space telescopes to explore the energy fractionation between the NUV and optical bandpasses. 

In \citet{Brasseur_2019ApJ...883...88B} we reported a previously uncatalogued population of $\sim$2,000 mostly small, short duration flares found in light curves from the GALEX space telescope. Because the stellar population we drew on was stars targeted by both the Kepler and GALEX missions, the majority of the targets are G-type main sequence stars. Our analysis indicated that the stars we did and did not detect flares on had consistent activity indicators, meaning we saw flares on even generally inactive stars. We found that the cumulative energy distribution of the body of GALEX flares followed a power law that agreed with other stellar and solar flare calculations, across a variety of wavebands. Comparing the duration-energy distribution of the GALEX flare population versus the Kepler flare populations from \citet{Namekata_2017ApJ...851...91N}, showed a lack of dependence between flare duration and energy, with a spread of three orders of magnitude in duration with all stellar flare energies of approximately the same range.

The paper is organized as follows: \S \ref{sec:data} describes the data reduction, \S \ref{sec:analysis} describes the data analysis, \S \ref{sec:modeling} describes the results of new modelling approaches to understanding the NUV and white light flare flare observations, \S \ref{sec:discussion} discusses the findings and implications, and \S \ref{sec:conclusion} concludes.

\section{Data Reduction} \label{sec:data}

This work is a multiwavelength study using data from two space telescopes; the Galaxy Evolution Explorer (GALEX), an orbiting ultraviolet mission active between 2003 and 2013, and Kepler, an earth trailing optical mission active between 2009 and 2018. Kepler observed in the wavelength range 4300–8900 \AA, with a spatial resolution of 4 arcseconds covering 105 deg$^{2}$. During its main mission (2009-2013) Kepler continuously monitored a single field of view producing continuous 30-minute (long) cadence, and periodic 1-minute (short) cadence light curves for its selected targets \citep{Kepler_paper}. Between 2009 and 2013, when both missions were in operation, GALEX observed in the Kepler field of view multiple times. The GALEX detectors output time-tagged photon events with a time resolution of 5 thousandths of a second, spatial resolution of 4-6 arcseconds and a field of view of 1.1 deg$^{2}$. While the GALEX telescope was equipped with both far- and near-ultraviolet (FUV and NUV) detectors, during the period of overlap with Kepler, only the NUV (1771-2831 \AA) detector was operating. The GALEX light curves in this study were produced with the software package gPhoton \citep{gphoton_ref} at a 10 second cadence, reduced and analyzed as described in \S 2 of \citet{Brasseur_2019ApJ...883...88B}.

Our entire analysis pipeline, including code for generating the figures in this paper, is available on GitHub \footnote{\url{https://github.com/ceb8/optical\_nuv\_multiwavelength\_flare\_study}}.

\subsection{Data Cross-matching} \label{sec:xmatch}

As discussed in \citet{Brasseur_2019ApJ...883...88B} we considered all GALEX data obtained within the Kepler main mission time frame without regards to specific quarters or long cadence vs short cadence coverage. Table \ref{tbl:overlaps} details the overlaps between GALEX and Kepler data and flares previously identified in these datasets. \citet{Brasseur_2019ApJ...883...88B} started with 34,271  stars seen by both Kepler and GALEX at some point during their four year period of overlapping observations, with  1,904 NUV flare detections (Figure 8 of \citet{Brasseur_2019ApJ...883...88B} shows a histogram comparing flaring and non-flaring stars). When considering strictly simultaneous overlap with Kepler long cadence data, these numbers reduce to 32,056 stars and 1,557 flares, respectively. Because of the smaller number of targets selected for Kepler short cadence observations, the overlap between the GALEX NUV flare detections and the Kepler short cadence data was very small, with 270 stars having at least some simultaneous overlap, and only two GALEX flares on a single star having simultaneous Kepler short cadence data. We also crossmatched the GALEX light curves against the \citet{Balona_2015MNRAS.447.2714B} Kepler short cadence flare catalog, and \citet{Yang_2019ApJS..241...29Y} long cadence flare catalog, which resulted in no Kepler short cadence flares and 7 long cadence flares with simultaneous GALEX data. These catalogs did not perform additional filtering on top of flare detections (such as requiring a minimum number of flares to be detected per star) and so were more suitable to our exploration of parameter space than e.g. \citet{Davenport_2016ApJ...829...23D}. Additionally, beyond the simultaneous data overlaps, twelve stars appeared in both the GALEX flare catalog \citep{Brasseur_2019ApJ...883...88B} and the \citet{Yang_2019ApJS..241...29Y} flare catalog, but without  simultaneous data for any of the flares.

\begin{deluxetable}{cc|c|c|}
  \tablewidth{0pt}
  \tablecolumns{4}
  \tablecaption{Temporal overlap of GALEX vs Kepler data (light curves) and flares. 
    \label{tbl:overlaps}}
  \tablehead{& & \multicolumn{2}{c|}{GALEX}\\\cline{3-4} Kepler & & Flares\tablenotemark{**} & \# of stars\tablenotemark{$\dagger$}}
  \startdata
\multicolumn{1}{ |c  }{\multirow{2}{*}{Short Cadence} } &
\multicolumn{1}{ |c| }{Flares\tablenotemark{*}} & 0 & 0       \\ \cline{2-4}
\multicolumn{1}{ |c  }{}                        &
\multicolumn{1}{ |c| }{\# of stars\tablenotemark{$\dagger$}} & 2 &   270     \\ \cline{1-4}
\multicolumn{1}{ |c  }{\multirow{2}{*}{Long Cadence} } &
\multicolumn{1}{ |c| }{Flares\tablenotemark{$\ddag$}} & 0 & 7  \\ \cline{2-4}
\multicolumn{1}{ |c  }{}                        &
\multicolumn{1}{ |c| }{\# of stars\tablenotemark{$\dagger$}} & 1557 & 32056  \\ \cline{1-4}
    \enddata
  \tablenotetext{*}{Taken from flare compilation of
  \citet{Balona_2015MNRAS.447.2714B}.}
  \tablenotetext{\dagger}{Temporal overlap deduced after positional cross-matching.} 
  \tablenotetext{$\ddag$}{Taken from flare compilation of \citet{Yang_2019ApJS..241...29Y}.} 
  \tablenotetext{**}{Taken from flare compilation of \citet{Brasseur_2019ApJ...883...88B}.} 
\end{deluxetable}

\subsection{Kepler Light Curve Detrending}\label{sec:detrending}

Here we consider the Kepler data which overlaps previously identified GALEX flares, as noted in Table~\ref{tbl:overlaps}. Additional flare detection is needed beyond that presented in the Kepler flare catalogs of \citet{Balona_2015MNRAS.447.2714B} and \citet{Yang_2019ApJS..241...29Y} to enable constraints appropriate to the strictly simultaneous overlap. We performed detrending of both short and long cadence light curves (irrespective of data quality flags) prior to performing flare searches. The goal of the detrending was to remove any periodic light curve variability (e.g. due to rotation) as well as instrument systematics. For the short cadence (1 min) Kepler data sigma clipping at 5$\sigma$ significance removed instrumental artifacts. Median detrending to remove longer timescale photometric variations followed. The window width depended on the dominant period in the light curve as determined from Lomb-Scargle periodogram analyses; the frequency cuts were determined empirically as follows:
\begin{eqnarray}
freq < 1~min^{-1},\textrm{ window = 401 min} \nonumber\\
1~min^{-1} \leq freq < 3~min^{-1},\textrm{ window = 201 min} \\
freq \geq 3~min^{-1},\textrm{ window = 101 min}\nonumber
\end{eqnarray}

For the long cadence Kepler light curves, we adapted the detrending method from \citet{Davenport_2016ApJ...829...23D}. Before detrending, we split each light curve into continuous sections with breaks of no more than 1 hour. Our detrending methodology is as follows:
\begin{enumerate}
    \item Two-pass multi-boxcar\\
    For each pass this consists of rolling median smoothing (minimum kernel size of 4), followed by sigma clipping ($\sigma=5$) and interpolation to fill in the rejected points.
    
    \item Five-pass fit sine curve \\
    For each pass, a Lomb Scargle periodogram is calculated and if the highest power period has power $> 0.25$, fit a sine curve with that period and subtract it from the flux.\\
    The sum of the five fit sine curve is also saved as the ``model.''
    
    \item Three-pass multi-boxcar\\
    Each pass is the same as in step 1.
    
    \item Twenty-pass Iterative Re-weight Least Squares (IRLS) spline fit\\
    Each pass first tries SciPy's LSQUnivariateSpline function\footnote{\url{https://docs.scipy.org/doc/scipy/reference/generated/scipy.interpolate.LSQUnivariateSpline.html}}, and  if that fails, falls through to SciPy's UnivariateSpline function\footnote{\url{https://docs.scipy.org/doc/scipy/reference/generated/scipy.interpolate.UnivariateSpline.html}}.
    
    \item Sum the result of step 2 to the sine model obtained in step 4 to determine the final quiescent flux model.
\end{enumerate}
 
After detrending both short and long cadence light curves we applied four different goodness-of-fit tests to determine the quality of our detrending. We considered 2 or more failures amongst the four tests as an overall failure and excluded those light curves from further exploration. We decided on this method by running a number of statistical tests on a selection of detrended light curves that had been hand-marked as well or poorly detrended. From these results we determined which tests were most effective for distinguishing good from poor detrending and that a single test was not sufficient for making the distinction without a great deal of false positives. The goodness-of-fit tests (all performed on a 1000 bin histogram of the detrended fluxes) and thresholds for success/failure we settled on are as follows:

\begin{itemize}
    \item Chi-squared test\\
    We calculated the chi-squared statistic between the histogram of the detrended fluxes and a Gaussian distribution constructed by setting the standard deviation to the 68\% width of the histogram, and the center to 0, and then normalizing it to have the same maximum as the detrended flux histogram.\\
    If the chi-squared statistic is $>$ 5000, the test fails.

    \item Jarque-Bera test \citep{Jarque_Bera}\\
    This is a statistical test of whether a distribution's skewness and kurtosis match that of a normal distribution. We use the SciPy implementation, and run it on the detrended flux histogram.\\
    If the Jarque-Bera statistic is $>$ 20000, the test fails.
    
    \item 90\% width test\\
    We calculate the range within which 90\% of the detrended (relative) flux values fall.\\
    If the 90\% width is $>$ 0.02, the test fails.
    
    \item Number of histogram peaks\\
    The detrended flux histogram should be a smooth gaussian shape, however for certain failure modes there are distinct peaks to either side of the central one. Because the calculated histogram is not an idealized gaussian, we performed 51-bin median smoothing before calculating the number of peaks.\\
    If the number of peaks is $>$ 1 the test fails.
\end{itemize}

\begin{figure}[!h]
    \includegraphics[width=0.5\textwidth]{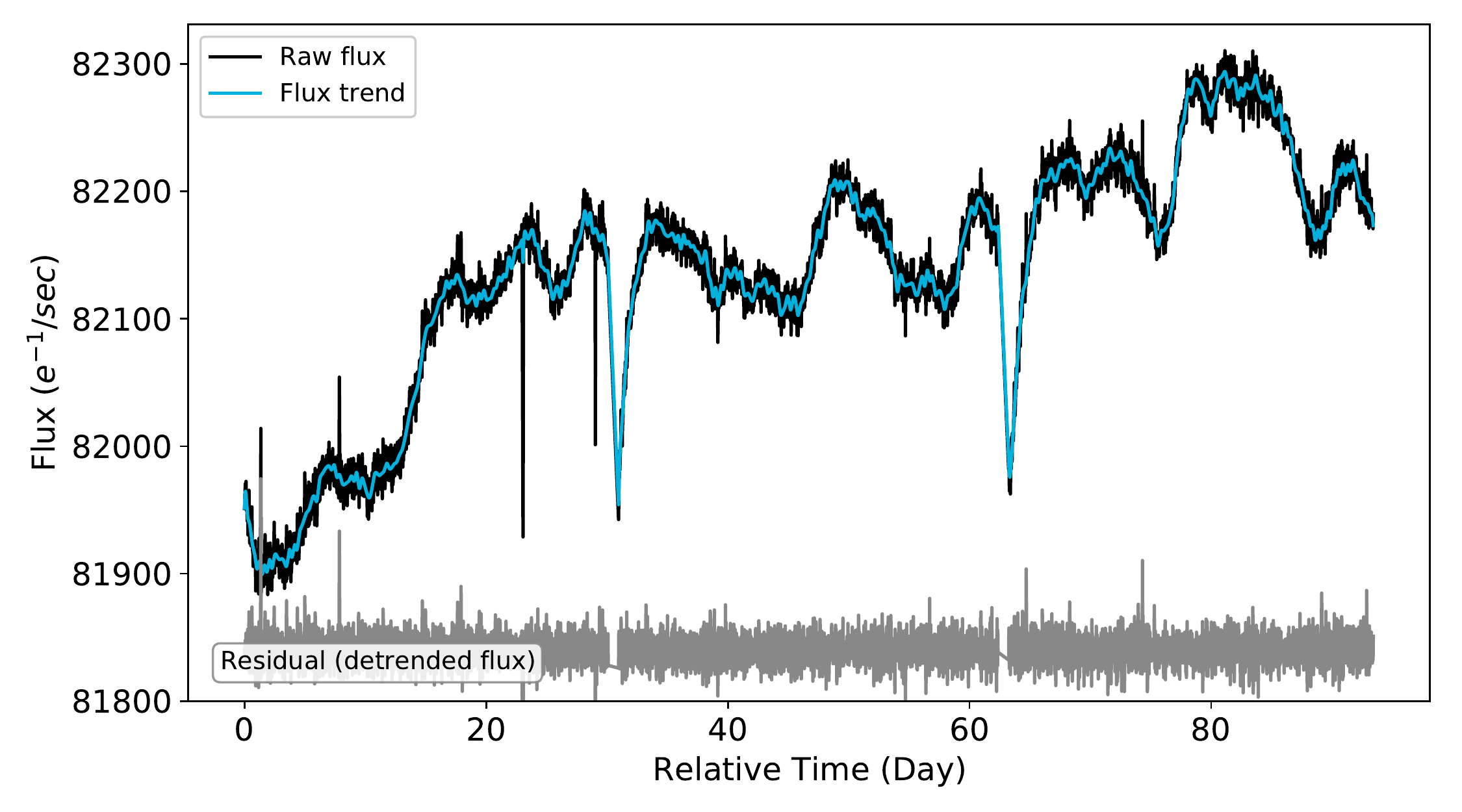}
    \includegraphics[width=0.5\textwidth]{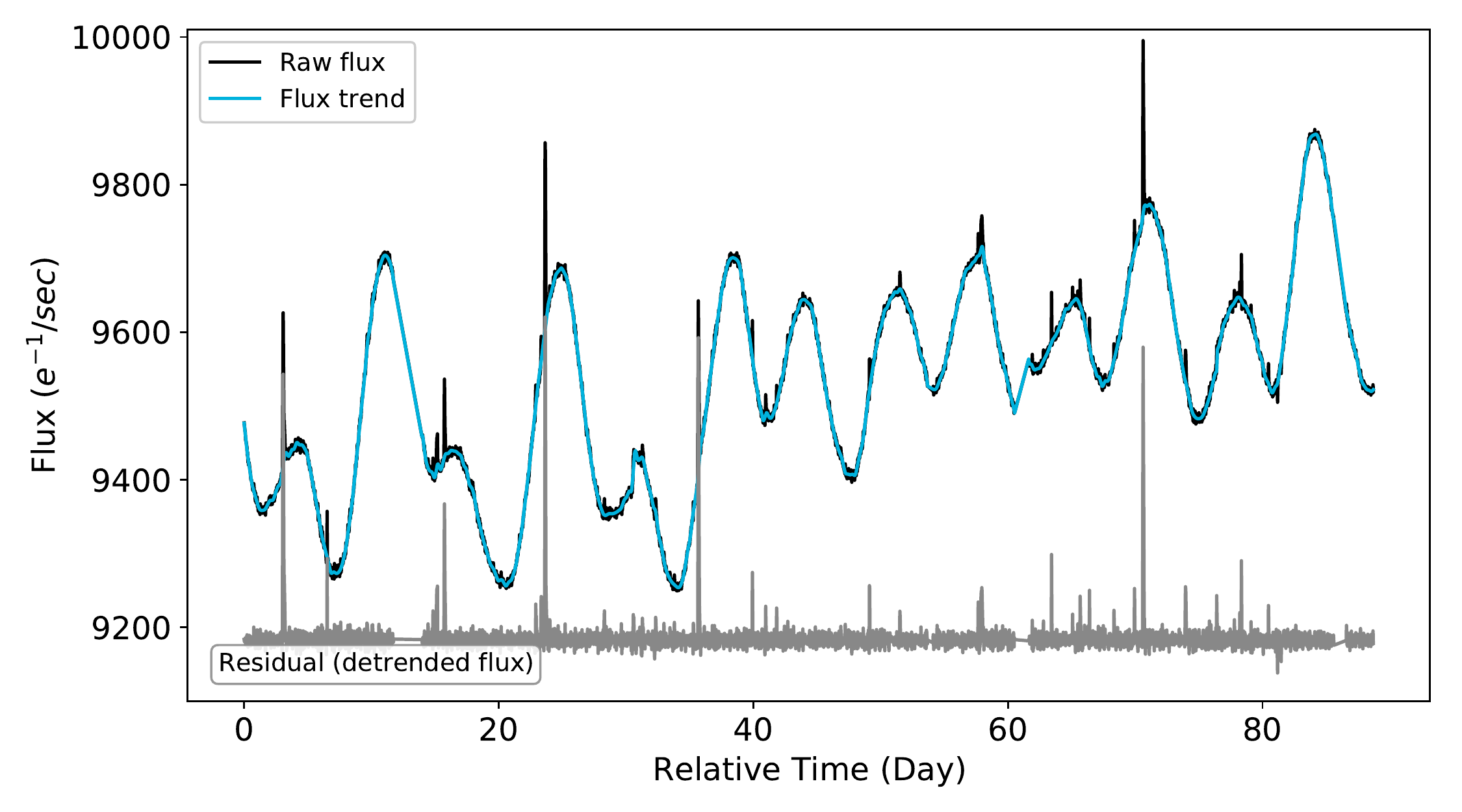}
    \caption{Here we show two well detrended light curves, one without flares (KIC 7350626 \textit{left}), and one with flares (KIC 9573767 \textit{right}).
    Black indicates raw flux and blue delineates the best-fit model as determined in steps 1-5 of \S\ref{sec:detrending}. The gray curve shows the residual (detrended) flux.
    \label{fig:detrending1} }
\end{figure}

\begin{figure}[!h]
    \includegraphics[width=0.6\textwidth]{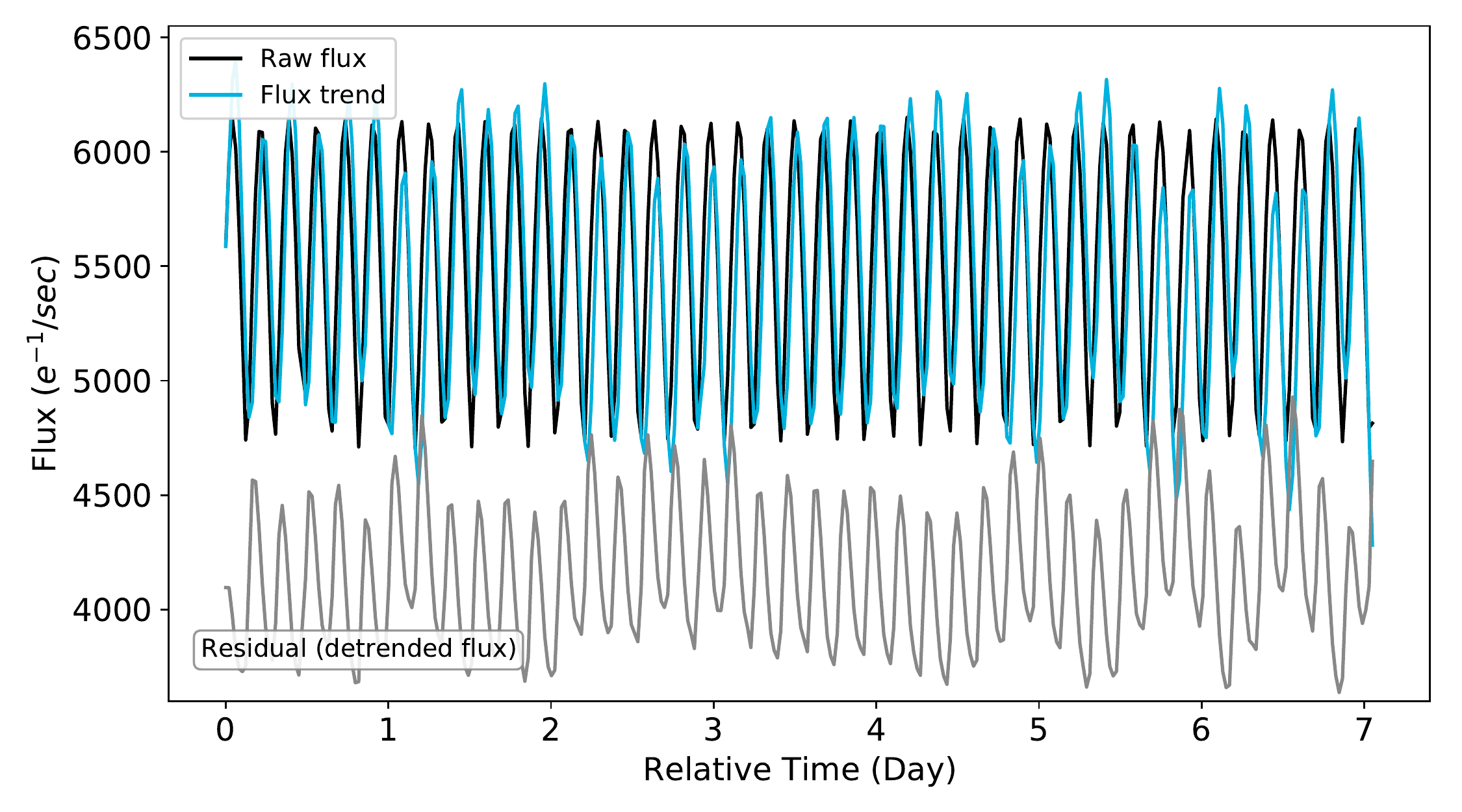}
    \includegraphics[width=0.325\textwidth]{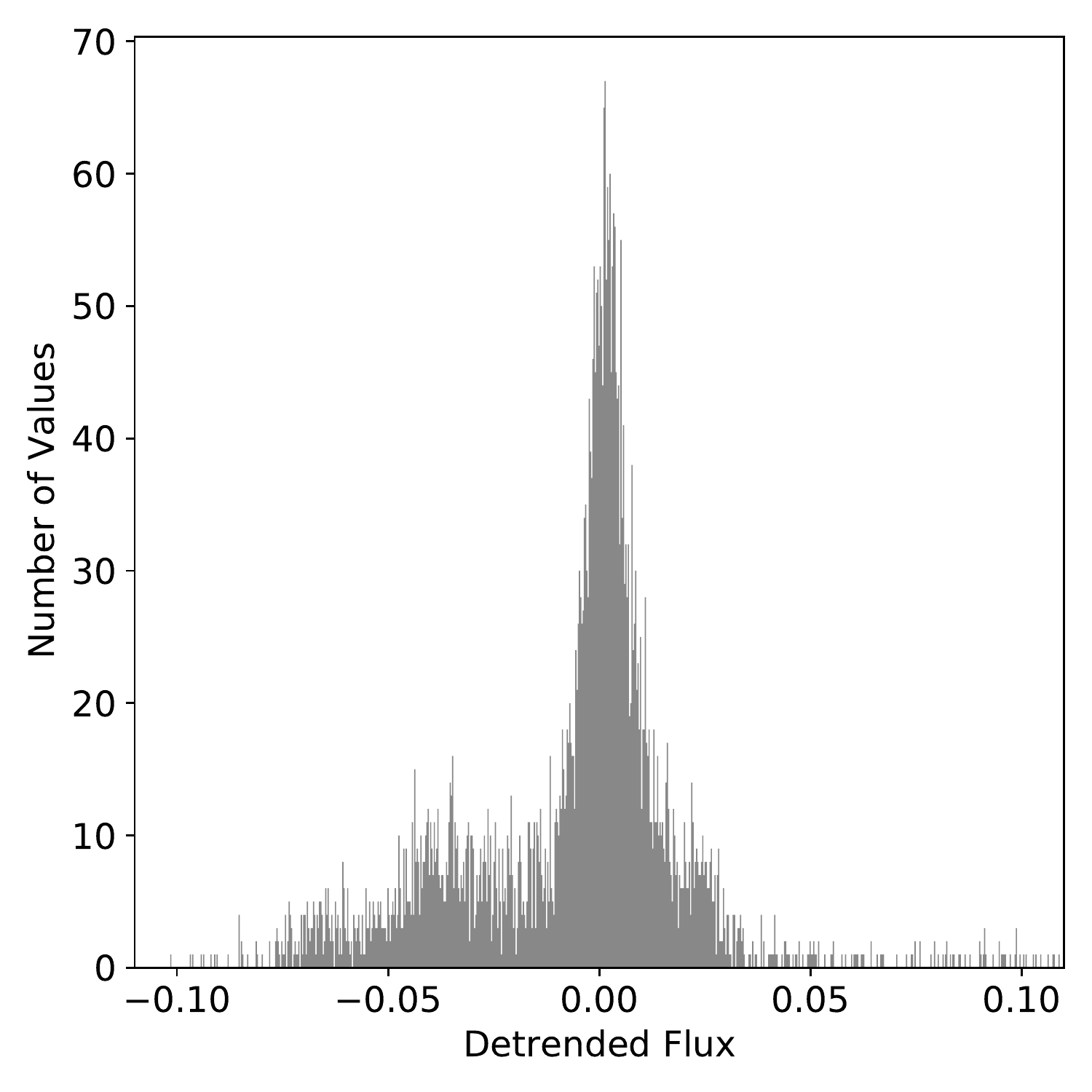}
    \caption{This figure shows how a light curve that has large-scale stellar photometric variations on similar timescale to flaring activity can fail the detrending process. The detrended light curve (KIC 10020247) is shown on the \textit{left} and its detrended flux distribution on the \textit{right}.
    The  histogram of flux values relative to the median shows a secondary peak to the left of the primary peak, leading the light curve to fail the number of peaks test and the 90\% width test.
    Black indicates raw flux and blue delineates the best-fit
    model as determined in steps 1-5 of \S\ref{sec:detrending}. The gray curve shows the residual (detrended) flux.
    \label{fig:detrending2} }
\end{figure}

Figure \ref{fig:detrending1} shows examples of two well detrended light curves, while Figure \ref{fig:detrending2} shows a poorly detrended light curve, and the histogram of flux values relative to the median. Only four long cadence light curves failed our detrending tests, so we simply discarded them.

\FloatBarrier

\subsection{Kepler flare detection} \label{sec:flarefinding}

We use the flare detection method outlined in \citet{Osten_2012ApJ...754....4O}, which was in turn based on \citet{Welch_1993AJ....105.1813W} and \citet{Stetson_1996PASP..108..851S}. We applied this method to the detrended light curves. 
This method uses  a statistic 
\begin{equation}
  \phi_{VV} = \left( \frac{V_{rel}}{\sigma} \right)_{i} \times \left( \frac{V_{rel}}{\sigma} \right)_{i+1}  
\end{equation}
where $V_{rel,i}$ is the relative (detrended) flux at cadence $i$ and $\sigma_i$ is the error in the flux measurement. This statistic is calculated for every temporal pair of values ($i$ and $i+1$ for all $i$). Each $\phi_{VV}$ pair is labeled as null ($\phi_{VV} < 0$), flare candidate ($\phi_{VV} > 0$, $V_{rel,i,i+1}>0$), and excluded ($\phi_{VV} > 0$, $V_{rel,i,i+1}<0$). We then calculated the histogram for the absolute value of the null sample, and the candidate sample. We fit the absolute value of the null samples with a double exponential of the form
\begin{equation}
\alpha e^{-\frac{x-a}{b}} + \beta e^{-\frac{x-a}{c}}  
\end{equation}

\begin{figure}[!h]
 \centering
    \includegraphics[width=\textwidth]{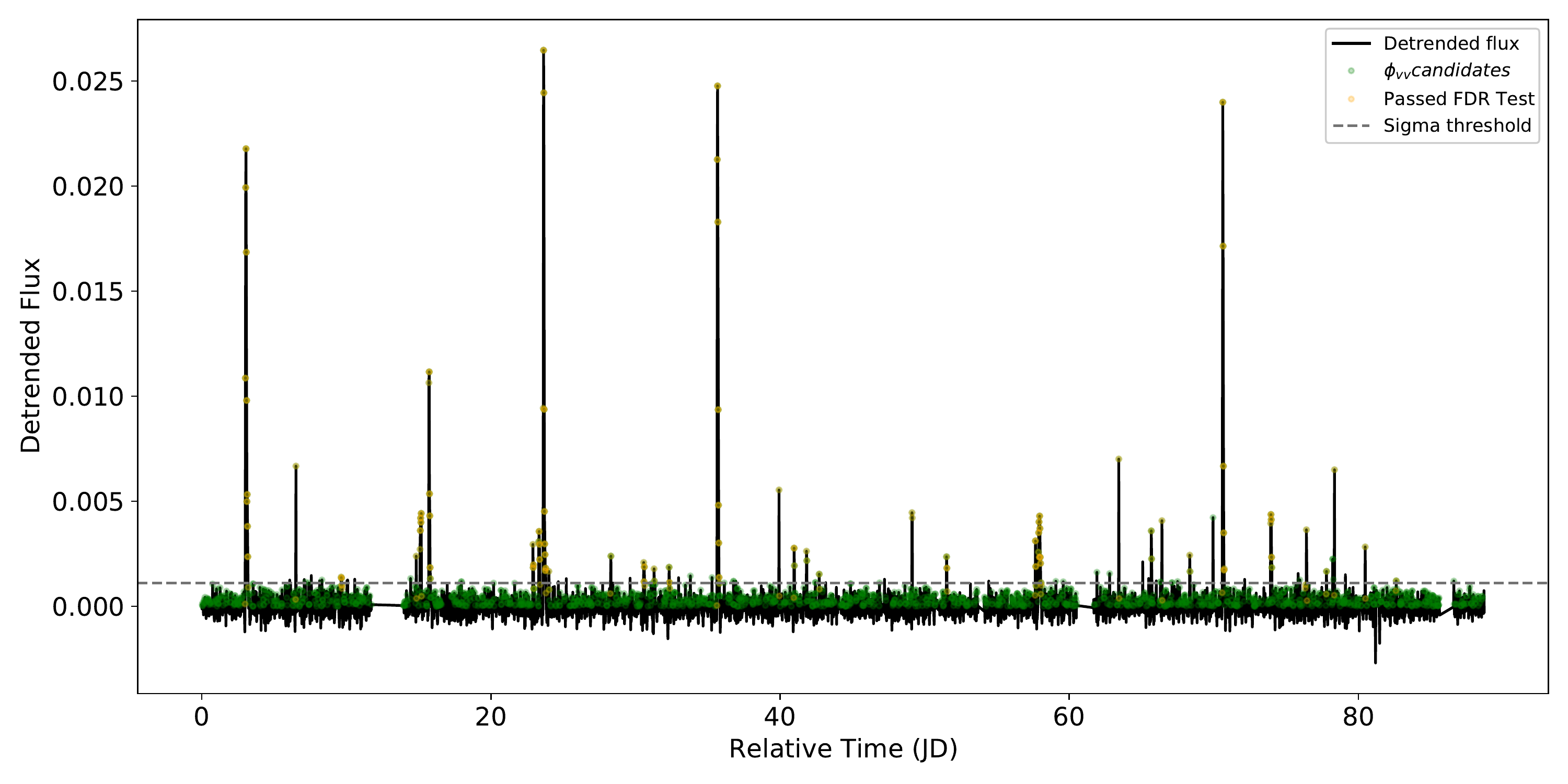}
    \caption{The flare detection process used in this paper, as applied to KIC 9573767 for Kepler quarter Q2. Black points show the detrended flux variations after removal of long timescale instrumental and stellar effects, as described in \S\ref{sec:detrending}. Green points indicate candidate flare data points based on $\phi_{VV}$ values. Yellow points delineate those remaining after filtering based on the False Discovery Rate. The dashed line indicates the location at 2.5$\sigma$ above the quiescent flux which sets a floor for candidate flare points. 
    \label{fig:flarefinding} }
\end{figure}

We next calculate the p-value for each candidate $\phi_{VV}$, where the p-value is the probability that a $\phi_{VV}$ from the (absolute value) null distribution has a value equal to or greater than the candidate $\phi_{VV}$ in question. Like \citet{Osten_2012ApJ...754....4O}, we use false discovery rate (FDR) analysis as described in \citet{Miller_2001AJ....122.3492M} to determine a threshold used to limit candidate points to those with $\phi_{VV}$ values at or above that threshold. We set the false discovery rate to 0.1 or 10\%. Additionally we apply a flux threshold of $2.5\sigma$ so that no points below $2.5\sigma$ above the quiescent flux are considered flaring. Figure \ref{fig:flarefinding} shows an example light curve with candidate points marked in green, points passing the FDR analysis in yellow, and the $\sigma$ threshold marked with the dotted gray line.

Once we have our final list of flare candidate points that pass FDR analysis and the $\sigma$ threshold, we turn it into a list of flares by finding flare edges for each point or group of points marked as flaring. Flare edges are considered to be the first point that falls below the quiescent flux to either side of the starting point(s). We also discarded flares within 4 points of a light curve section edge. This is because our detrending algorithm is not as good at the edges so flares found there are considered suspect. See \S\ref{sec:injection} for a discussion of the false positive rate with this flare detection technique.

\FloatBarrier

\subsection{Kepler flare energy calculation}\label{sec:energy}

When determining the energy of detected flares in Kepler light curves, we calculated a local quiescent flux line and subtracted it from the raw light curve to obtain the flare-only fluxes. We did this instead of using the detrended light curves because for larger flares the light curve detrending also removed some of the flare flux along with the larger light curve trends we were intentionally smoothing away. To calculate the local quiescent flux for an individual flare we fit a line to 10 points on either side of the flare, with a 5 point buffer between the flare and the presumed quiescent flux points. The buffer was to account for cases where the calculated flare edges were incorrect. After calculating the local quiescent flux, we used that line and the raw light curve to adjust the flare bounds. This step occasionally turned two flare detections into a single detection, so we removed the duplicates. 

Once this list of flare photometry points and quiescent values was assembled we converted the Kepler flux values (e$^-$ s$^{-1}$) to physical flux values (erg s$^{-1}$ cm$^{-2}$) using the Kepler zero point, bandpass central wavelength and full width at half maximum (FWHM) \citep{Kepler_bandpass_ref}. The flare energies were then calculated using the method described in \S3.2.3 of \citet{Brasseur_2019ApJ...883...88B}, stopping short of converting the bandpass-specific energy to bolometric energy (as that is a topic of investigation of the current paper). Note that standard formulae for reddening and extinction have been applied, with bandpass specific  extinction values given by \citet{Yuan_etal_2013}, and individual reddening values calculated with the dustmaps package \citep{dustmaps}.

\FloatBarrier

\section{Results and Analysis} \label{sec:analysis}

\subsection{Detection of Flares in Kepler Data at the Times of GALEX Flares}

The flare detection methods applied to the Kepler data during the time of overlap with GALEX observations did not return any detectable flares. The flare detection algorithms can also be used to determine sensitivity to flares of different sizes via synthetic flare injection. This is important for placing limits on the sizes of flares which may be occurring but undetectable due to observing constraints.

\subsection{Synthetic Flare Injection and Subsequent Recovery} \label{sec:injection}

Synthetic flare injection and recovery aided in determining the parameter space for which flares are detectable in Kepler long cadence data. Because there is so much more Kepler data than GALEX, it was not possible to perform manual verification on all automatically detected flares; instead we injected a known sample of synthetic flares, and tested our flare detection and characterization algorithms against that sample.

To create the injected flares, we used the model flare equations from \citet{Davenport_2014ApJ...797..122D}:

\begin{equation}
F_{rise} = 1+1.941(\pm 0.008) t_{1/2} - 0.175 (\pm 0.032) t_{1/2}^2 - 2.246 (\pm 0.039) t_{1/2}^3 - 1.125(\pm 0.016) t_{1/2}^4, \; -1 \le t_{1/2} \le 0
\end{equation}

\begin{equation}
F_{decay} = 0.6890(\pm 0.0008) e^{-1.600(\pm 0.003)t_{1/2}} + 0.3030(\pm 0.0009)e^{-0.2783(\pm 0.0007)t_{1/2}}, \; t_{1/2} \ge 0
\end{equation}
The flux values are normalized as ($F = \frac{(f-f_{qui})}{(f_{peak}-f_{qui})}$) and time normalized to the FWHM ($t_{1/2} = \frac{t-t_{peak}}{FWHM}$). Our method for flare injection was to first choose a flare energy and $t_{1/2}$, then combine these values with the distance measurement  from \citet{Bailer-Jones2018AJ....156...58B}, which allowed us to determine the maximum flux for the flare on a given star. The benchmark photo-electron current at the Kepler focal plane for a 12th magnitude star \citep{Kepler_2016ksci.rept....1V} then enabled a scale to transform this physical flux into a Kepler flux with units of $e^-$ s$^{-1}$. 

Injected flares follow the distribution of flare occurrence rate with energy found in the population of flares detected in the NUV GALEX data in \citet{Brasseur_2019ApJ...883...88B}. 
The probability distribution function is 

\begin{equation}
 P(E)=(\frac{E}{E_{min}})^{1-\alpha}   
\end{equation}
with best fit parameters $E_{min}=7.2\times 10^{34}$ erg, and $\alpha = 1.72 \pm$ 0.05. We used inverse transform sampling to sample this distribution

\begin{equation} 
\label{eq:probdist}
P^{-1}(u) = E_{min}u^{\frac{1}{1-\alpha}}
\end{equation}

For injection we used the best fit $\alpha = 1.72$ from \citet{Brasseur_2019ApJ...883...88B}, but the minimum energy of injected flares was set to a much lower number, $E_{min}=3\times 10^{31}$ erg, in order to sample the entire range of detected GALEX flares. Flare injection into the raw (not detrended) Kepler light curves proceeded at a rate of $\sim 10$ flares per day (0.4 per hour), with flare energies drawn from the probability distribution given in equation \ref{eq:probdist}, and the FWHM drawn from a uniform probability distribution between 30 seconds and 1 hour. Flares were injected irrespective of flares already existing in the light curve. However, flares recovered that overlapped existing flaring activity were removed from our sample to avoid contamination. Detrending and flare finding of these injected light curves proceeded in the same manner as described above for the original light curves (\S \ref{sec:detrending},\ref{sec:flarefinding}). To create the final list of recovered flares we cross-matched the list of flares detected in the injected light curve against both the original light curve (enabling the removal of detections of actual flaring activity) and the list of injected flares (the ``recovered flare'' was considered to be the most energetic injected flare peaking within the recovered peak flux time bin). The method outlined in \S \ref{sec:energy} was used to determine the energy of the recovered flares. 

\begin{figure}[!h]
 \centering
    \includegraphics[width=0.8\textwidth]{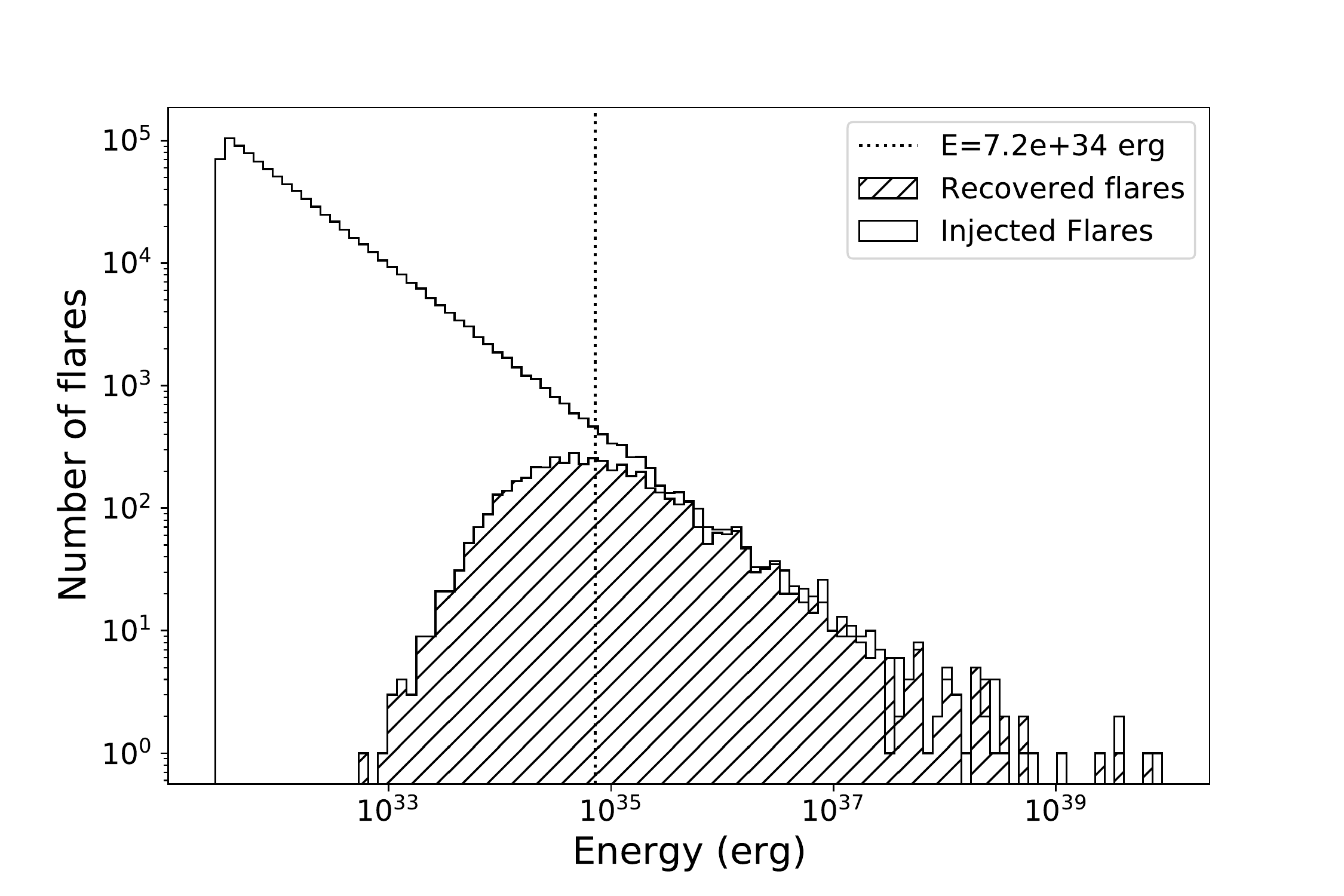}
    \caption{Injected vs. recovered flare energy histogram. Injected energies are the input energies in artificial flares, while recovered energies are the flares energies  recovered through flare detection algorithms. The dotted vertical line shows the minimum energy ($E_{\rm min}$) from \citet{Brasseur_2019ApJ...883...88B}. The energies on this plot are Kepler band only.
    \label{fig:injRecEngHist} 
}
\end{figure}

\begin{figure}[!h]
 \centering
    \includegraphics[width=0.8\textwidth]{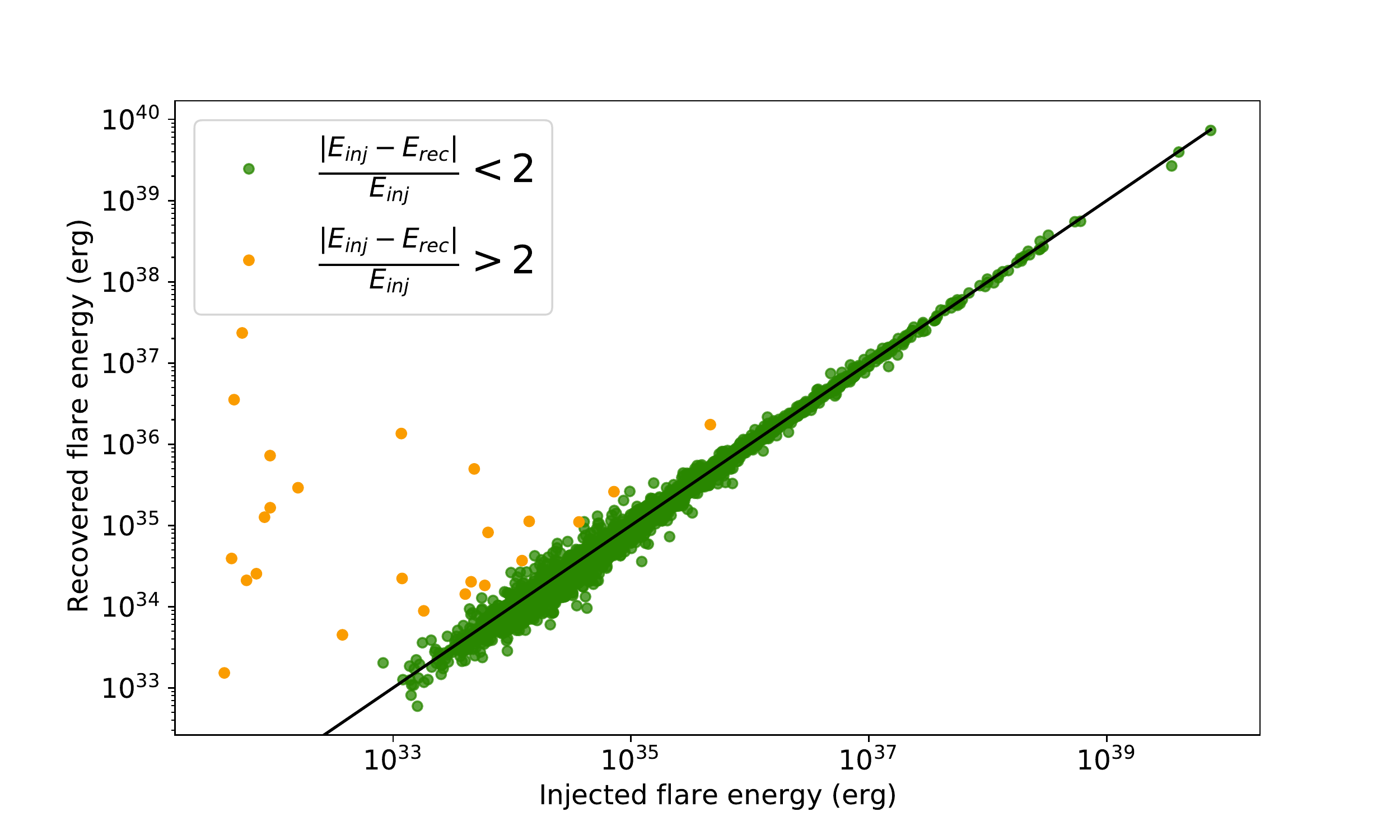}
    \caption{Scatter plot of the ability to recover flares of a given energy using the detrending and flare detection algorithms described in \S\ref{sec:detrending} and \S\ref{sec:flarefinding}. Points are color-coded according to how close the recovered energy matches the injected energy. Green points are those injected flares whose energies were recovered to within a factor of 2, while orange points are injected flares which were recovered with a much larger value of recovered energy. Because of their generally lower injected energy compared to the recovered flare energy, these are likely false detections that coincidentally occur near unrecoverable injected flares.
    \label{fig:injRecEngComp} }
\end{figure}

Figure \ref{fig:injRecEngHist} compares the injected and recovered flare distributions. The solid histogram indicates how many flares were injected into the light curves, while the hatched histogram displays those which were recovered after the detrending and flare detection methods described earlier. Approximately the same number of flares are injected and recovered starting at a few times $10^{35}$ ergs; below that value, there is a sharp turn off in the number of injected flares that are recovered. This turn-off indicates the minimum flare energy we are able to detect. It is not a hard cut-off due to the combined effects of false recoveries, and different stars having different minimum detectable energies. Figure \ref{fig:injRecEngComp} plots the injected (true) flare energy versus the recovered flare energy for individual flares which were both injected and recovered. Points colored in green have a recovered energy  within a factor of 2 of the injected energy. The injected and recovered energies are largely centered on the line of equality, with smaller deviations at higher energies. The largest outliers in recovered energy compared to injected energy are concentrated at low flare energies, and are thus most likely false identifications which happen to coincide with unrecoverable injected flares. Combining this result with Figure \ref{fig:injRecEngHist} reveals that below about $10^{34}-10^{35}$ ergs, most of the flare detections are false positives, even if the recovered energy agrees well with the injected energy. That there is not a hard cutoff below which all detections are false positives arises from two main considerations: (1) detectability depends on more than just the flare energy, and (2), these plots combine data from many stars at different distances, with distance being one of the main determinants of flare detectability as a function of energy.

At this point we added the additional condition that for a flare detection to be considered a true flare recovery, the injected flare energy must be ``near'' the recovered flare energy, where ``near'' is defined as $\frac{|E_{inj}-E_{rec}|}{E_{inj}} < 2$. This data cut removed $\sim1\%$ of the recovered flares from the sample.

\begin{figure}[!h]
 \centering
    \includegraphics[width=0.8\textwidth]{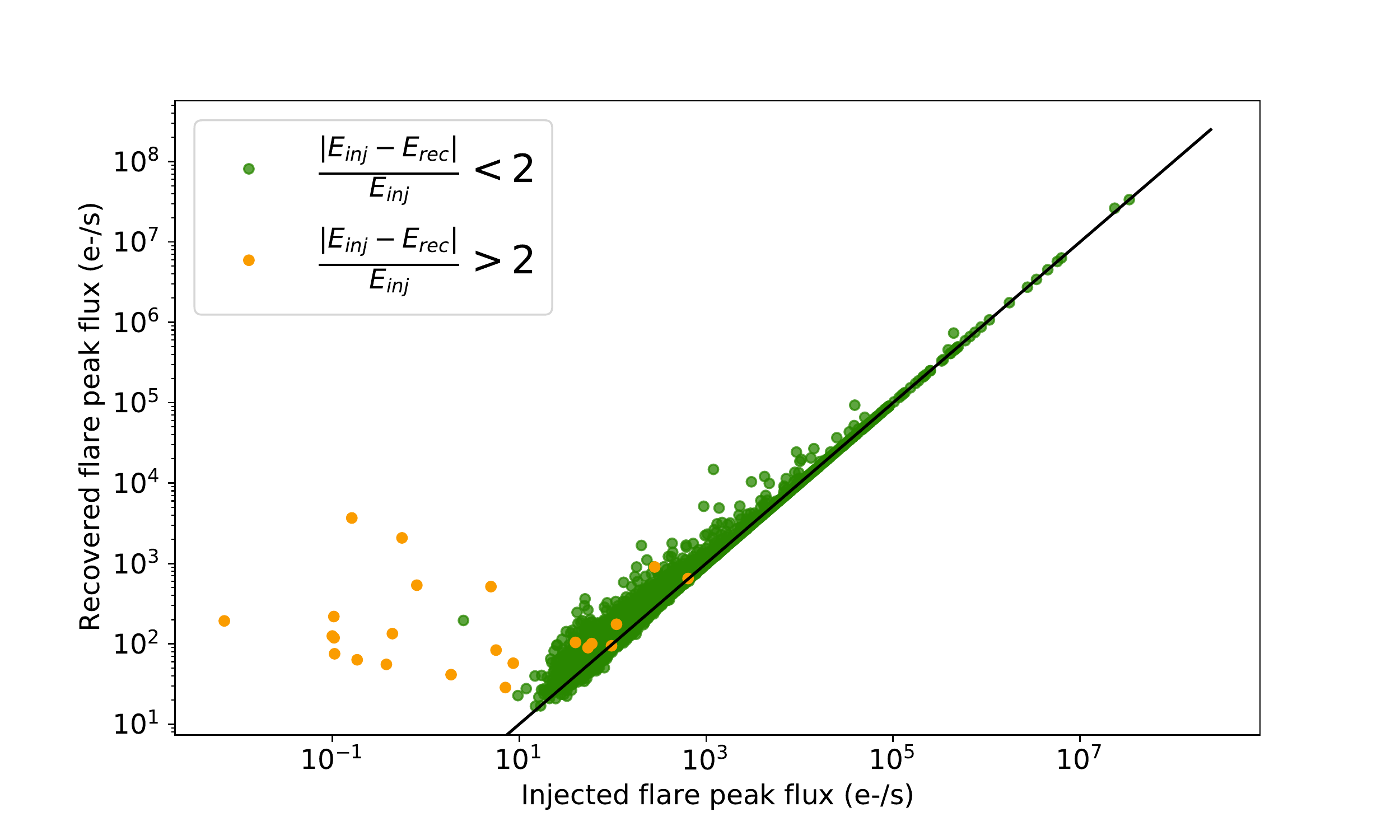}
    \caption{Scatter plot showing injected vs. recovered flare peak fluxes. The fluxes represent the maximum additional flux above quiescence due to the flare. Green points are those injected flares whose energies were recovered to within a factor of 2. Orange points are injected flares which were recovered, but without a good energy match (same color scheme as Fig. \ref{fig:injRecEngComp}). Successful recovery of the energy corresponds with successful recovery of the peak flux, based on the good agreement in recovering the energy and peak flux for the bulk of the sample, and the small number of outliers in both peak flux and energy. 
    \label{fig:injRecFluxComp} }
\end{figure}

Figure \ref{fig:injRecFluxComp} displays the same general pattern for injected flare peak flux vs recovered flare peak flux as is visible in the recovered vs injected energies of Figure~\ref{fig:injRecEngComp}. In this case the ``true recoveries'' are still defined by the recovered energy being close enough to the true energy, so there are a few false recoveries with peaks close to the true peak flux and vice versa. Notably only 2 true recoveries have peak fluxes more than 10 times larger than the injected flare peak flux, which indicates that when the injection and recovery method successfully recovers the energy it almost always also successfully recovers the peak flux. 

Overall our ability to recover flares is quite good above a certain threshold that varies per star, and our ability to recover flare bulk properties like energy distribution is even better. To quantify the range in which we can trust flare recoveries we calculated a minimum detectable flare energy and peak flux on a per star basis. We can then compare these minimum detectability thresholds with the population of GALEX flares.

To determine the minimum detectable energy for a given star, we calculate the percent of flares recovered above energy $E$ for $E$ from $10^{31}-10^{37}$ ergs as:
\begin{equation}
 P(E_i\in [0^{31},10^{37}]) = \frac{N(E_{rec}>E_i)}{N(E_{inj}>E_i)}
 \label{eqn:recovery}
\end{equation}
where $N(E_{rec}>E_i)$ is the number of flares recovered with energy greater than $E_i$ and $N(E_{inj}>E_i)$ is the number of flares injected with energy greater than $E_i$. We chose a threshold of 95\% of injected flares being recovered and $E_i$ with $P(E_i)$ nearest to the threshold as the minimum detectable energy. We used the same method to calculate the per star minimum detectable peak flux, independent from minimum detectable energy. Using the \citet{Davenport_2014ApJ...797..122D} model flare equations to relate peak flux, duration, and energy, we can produce a contour of possible flare peak flux/duration combinations  for a given minimum energy. 

\begin{figure}[!h]
 \centering
    \includegraphics[width=0.8\textwidth]{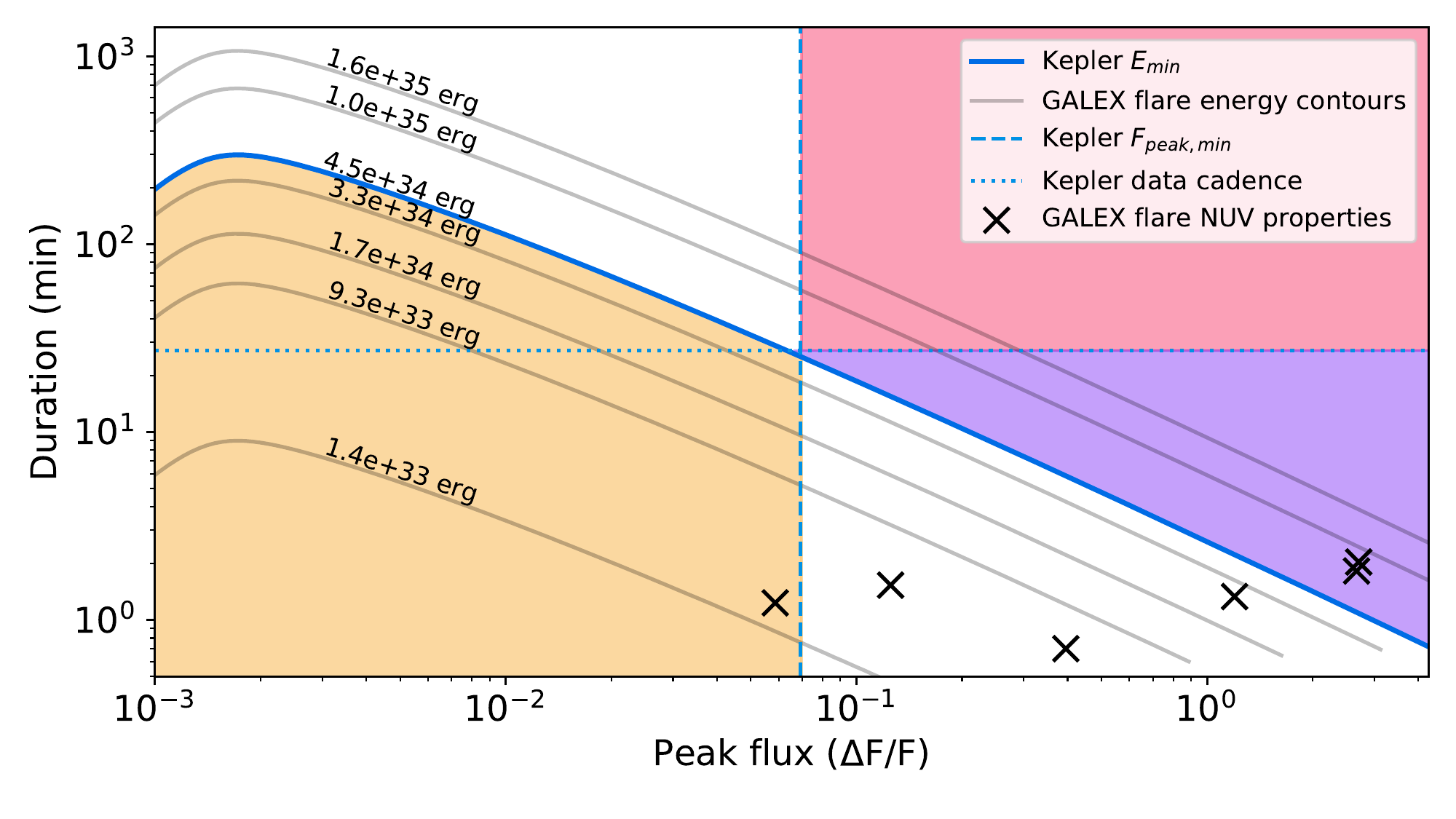}
    \caption{Contours of flare energy as a function of optical (Kepler-band) flux enhancement and flare duration for an example star, KIC 9775956. The solid blue contour is the minimum detectable energy in the Kepler bandpass as calculated through synthetic flare injection analysis (see Equation~\ref{eqn:recovery}). The gray contours are for individual flares detected in the KIC 9775956 GALEX light curve, using the energy fractionation of $E_{Kep}/E_{Gal} = 1.51$ calculated using Table 2 in \citet{Osten_Wolk_2015ApJ...809...79O} (see \S \ref{sec:discussion} for why this is not always a good estimate). The data points ($\times$) show the peak flux enhancement versus duration for the observed NUV GALEX-band flares. The values have not been converted into the Kepler band, but were rebinned to estimate the flux enhancement for a 30-minute rather than 10-second time bin. These data points do not fall on the associated gray contours both because they have not been converted from the NUV values, and because the GALEX flare morphologies do not conform to the \citet{Davenport_2014ApJ...797..122D} model which we use when calculating the energy contours. The horizontal dotted line indicates the Kepler long cadence of 30 minutes, and the vertical dashed line indicates the sensitivity limit of detectable fluxes. The red and purple shaded regions indicate where flares are detectable by both peak flux enhancement and energy measures, while the orange shaded region indicates where flares are not detectable by both measures. The purple region indicates flares that are detectable, but shorter in duration than the 30-minute data cadence as opposed to the red region where flares are detectable by energy/flux measure and should also span at least two data points.
    \label{fig:galex_kepler_contour} }
\end{figure}

Figure~\ref{fig:galex_kepler_contour} shows an example of this in practice: the peak flux vs. duration contours for KIC 9775956. The blue line is the peak flux/duration contour for KIC 9775956's minimum detectable energy. Here we plot the normalized flux enhancement on the x-axis, calculated as $\frac{F_{peak}-F_{qui}}{F_{qui}}$. The gray lines are the contours for the NUV flares identified in KIC 9775956's GALEX light curve, with the Kepler band energy estimated using the energy fractionation from \citet{Osten_Wolk_2015ApJ...809...79O} (which we assert is not a good estimate for all flares, see the discussion section). These contours are calculated by taking the observed energy in combination with the \citep{Davenport_2014ApJ...797..122D} flare model and plotting the curve of possible peak flux/duration values associated with the given energy. The minimum detectable peak flux for KIC 9775956 is marked (dashed vertical line) as is the Kepler data cadence (horizontal dotted line). The GALEX NUV peak flux and duration are also marked in the figure -- the peak flux values are the NUV GALEX-band fluxes, rebinned for the Kepler cadence by assuming the GALEX peak flux $\Delta F/F$ is the only enhancement in the Kepler bin, and extrapolating what that enhancement would be in a 30-minute rather than 10-second time bin. These points don't sit on their respective contours, both because they have not been converted from NUV to optical, and because the GALEX flares do not conform morphologically to the \citet{Davenport_2014ApJ...797..122D} model. 

This plot (Figure \ref{fig:galex_kepler_contour}) highlights the challenges in comparing data with such different time resolution. While the minimum detectable energy and peak flux values give a sense of flare detectability, these parameters do not present the entire picture due to observational constraints. The area shaded in red represents flares where the peak enhancement is greater than the minimum detectable peak flux, the energy is above the minimum detectable energy, and the duration is greater than the Kepler data cadence (30 minutes). These flares would be detectable. Likewise, the area shaded in orange represents flares where neither the peak flux nor energy meet the light curve's detectability thresholds; these flares are non-detectable. However, many of the the GALEX flares are in the region shaded in purple, where the flares are above both the peak enhancement and energy thresholds, but have durations shorter than a single Kepler cadence. Assuming the \citet{Osten_Wolk_2015ApJ...809...79O} energy fractionation, all of the GALEX flares have peak enhancements far above the minimum detectable peak flux, although not all of them are above the energy detection threshold. However, all of the GALEX flare durations are well below the Kepler data cadence, meaning that the entirety of the flare occurs within a single Kepler time bin. Thus, within the Kepler data, the flare flux enhancement will be diluted because the very short time duration of flux enhancement will be spread out across the data bin. Additionally it means we expect only a single flux measurement to be enhanced. As a result we expect that we cannot detect these flares directly in the Kepler data, but that given prior knowledge of an existing GALEX flare, that we may be able to see evidence of flux enhancement in the corresponding Kepler data point for the flares above the Kepler detection threshold.

\subsection{Kepler non-detections of GALEX flares} \label{sec:nondetections}

\subsubsection{Constraints from Short Cadence Kepler data}\label{sec:short_const}
\begin{figure}
    \centering
    \includegraphics[scale=0.5]{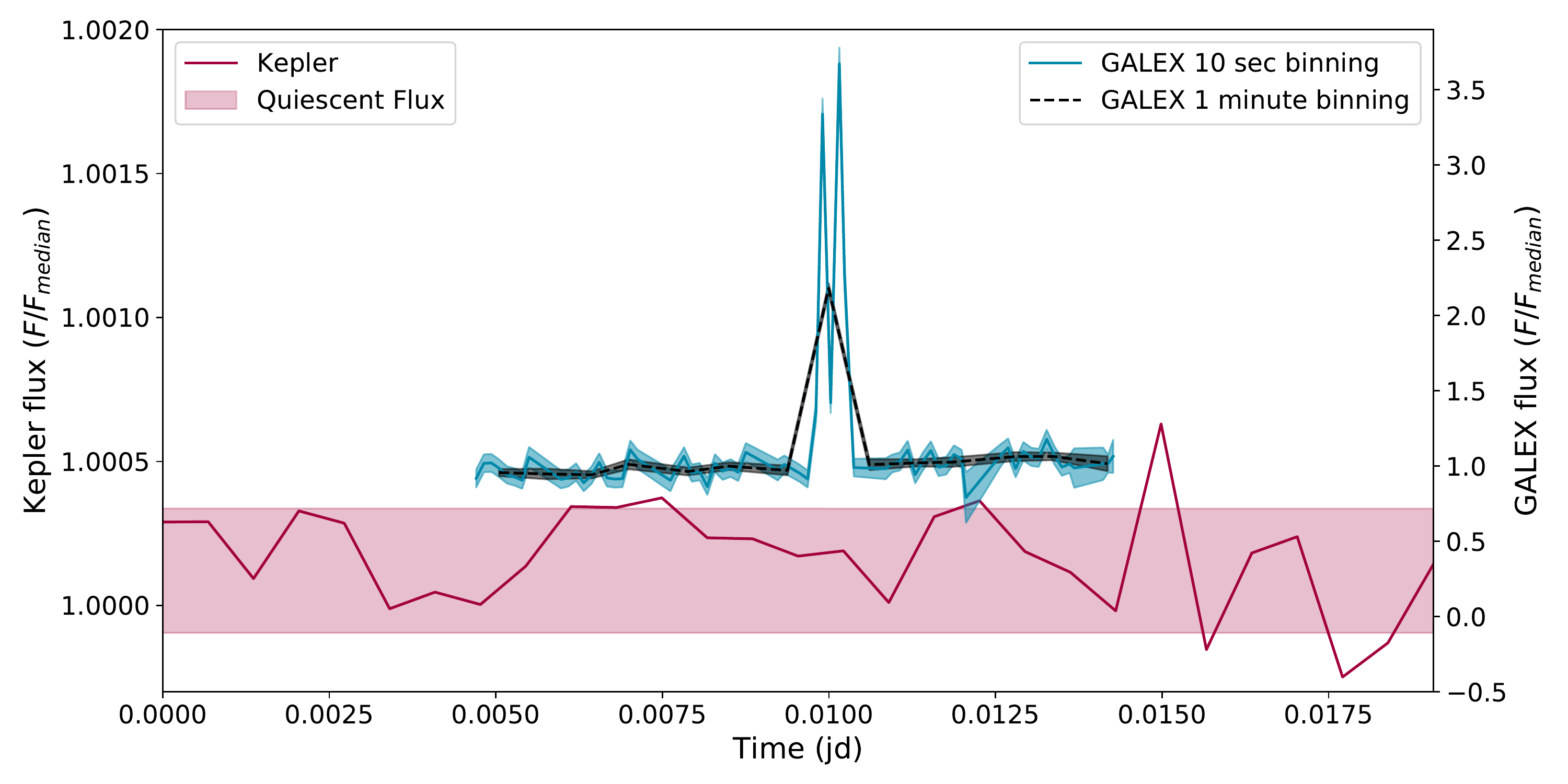}
    \includegraphics[scale=0.5]{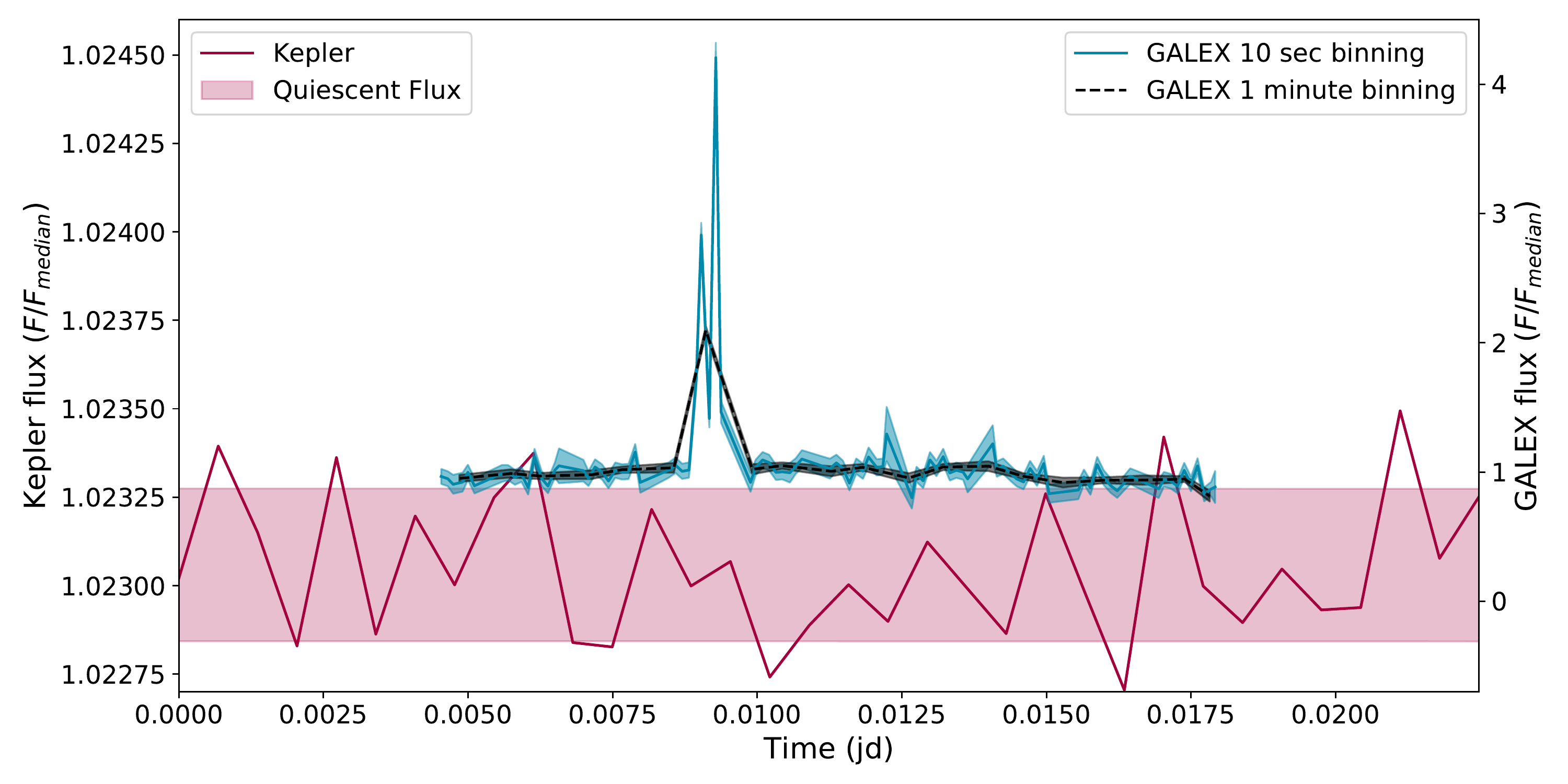}
    \caption{Plot showing the temporal overlap between two previously identified GALEX flares and simultaneous constraints from short cadence Kepler data: GK-1 (top) and GK-2 (bottom).  Both flares are on KIC 9592705, an eclipsing binary of F7V stars. The measured GALEX band flare energies and maximum energy Kepler flare that could be hiding in the Kepler light curve are shown in Table \ref{tbl:1minflares}. The blue lines are the GALEX light curve data at 10 second cadence, while the red line shows the Kepler 60 second light curve data. The black line is the GALEX data resampled at a 60 second cadence, an equivalent time sampling as the Kepler Short Cadence data. The pink region indicates the estimated 1 standard-deviation range of the quiescent flux in the region of the flare. Note the difference in scale between the GALEX relative flux scale (on the right hand side) and the Kepler relative flux scale (left hand side).
    \label{fig:GALEX_KeplerSC}}
\end{figure}

As discussed earlier (see Table \ref{tbl:overlaps}) only two of the detected GALEX flares had simultaneous Kepler short cadence data. This is the most advantageous cadence for comparison, as the one minute short cadence data is better matched to the ten second binning of the GALEX data. These two events are shown in Figure~\ref{fig:GALEX_KeplerSC}, and coincidentally occurred on the same star, KIC 9592705. We label the flares GK-1 (Figure~\ref{fig:GALEX_KeplerSC} \textit{top}) and GK-2 (Figure~\ref{fig:GALEX_KeplerSC} \textit{bottom}) and display the flare properties in Table \ref{tbl:1minflares}.

\begin{table}[]
    \begin{tabular}{|c|l|l|l|l|l|l|}
    \hline
 Flare             &F$_{\rm pk}/F_{\rm med}$ & F$_{\rm qu}/F_{\rm med}$ & E (erg) & F$_{\rm pk}/F_{\rm med}$  & F$_{\rm qu}/F_{\rm med}$ & E (erg) \\
 \cline{2-7} 
       & \multicolumn{3}{c|}{GALEX} & \multicolumn{3}{c|}{Kepler} \\
  
    \hline
     GK-1  & 2.18 & 0.99 &$9.94\times10^{33}$ & $<1.0004$ & 1.0001 & $<2.83\times 10^{32}$\\
     GK-2  & 2.10 & 0.99 & $9.36\times10^{33}$ & $<1.0235$ & $1.0231$ & $<1.91\times 10^{32}$ \\
    \hline
    \end{tabular}
    \caption{Flux and Energy Ratios for GALEX and Kepler short cadence flare data on KIC 9592705. Note that the GALEX values are based on the light curve binned to 1 minute cadence to match Kepler's short cadence.}
    \label{tbl:1minflares}
\end{table}

At the higher time resolution of the GALEX data, each NUV flare appears to contain two sub-bursts, and has a total duration on the order of a minute. The black curves display the GALEX data resampled to match the one minute cadence of the short cadence data. In this case the flux enhancement is limited to a single time bin, but the magnitude of the flare enhancement is still large at this cadence, at slightly more than twice the median flux.

In contrast, there is no apparent evidence of flux increases in the Kepler data at the time of the GALEX flares.  The red curve illustrates the Kepler flux expressed relative to the median Kepler flux.  The broad pink region indicates what the quiescent flux level is based on flux measurements to either side of the flare. For these two flares we can place an upper limit on the energetics of a white light flare which could be occurring  at the time of the NUV flare but remain undetected.

The star on which these GALEX flares occurred is KIC 9592705, described by Simbad as an eclipsing binary of F7V spectral type. \citet{zhang2019} examine binaries observed by Kepler and LAMOST, and return an orbital period of 10.27 days, with a primary T$_{\rm eff}$ of 6185 K and a distance of 409 pc, with lower and upper bounds of 405 and 414 pc respectively.

\FloatBarrier

\subsubsection{Constraints from Long Cadence Kepler data \label{sec:lc_con}}

\begin{figure}[!h]
 \centering
    \includegraphics[width=.9\textwidth]{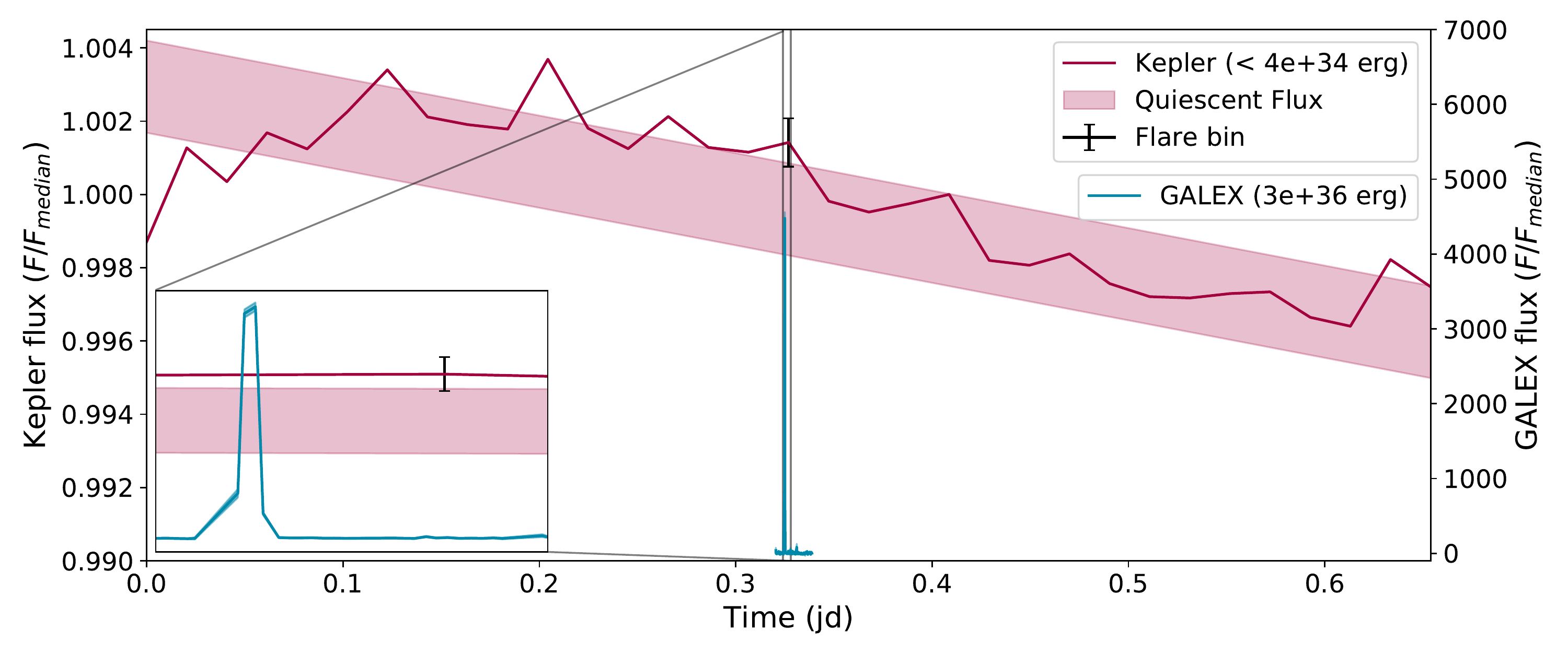}
    \includegraphics[width=.9\textwidth]{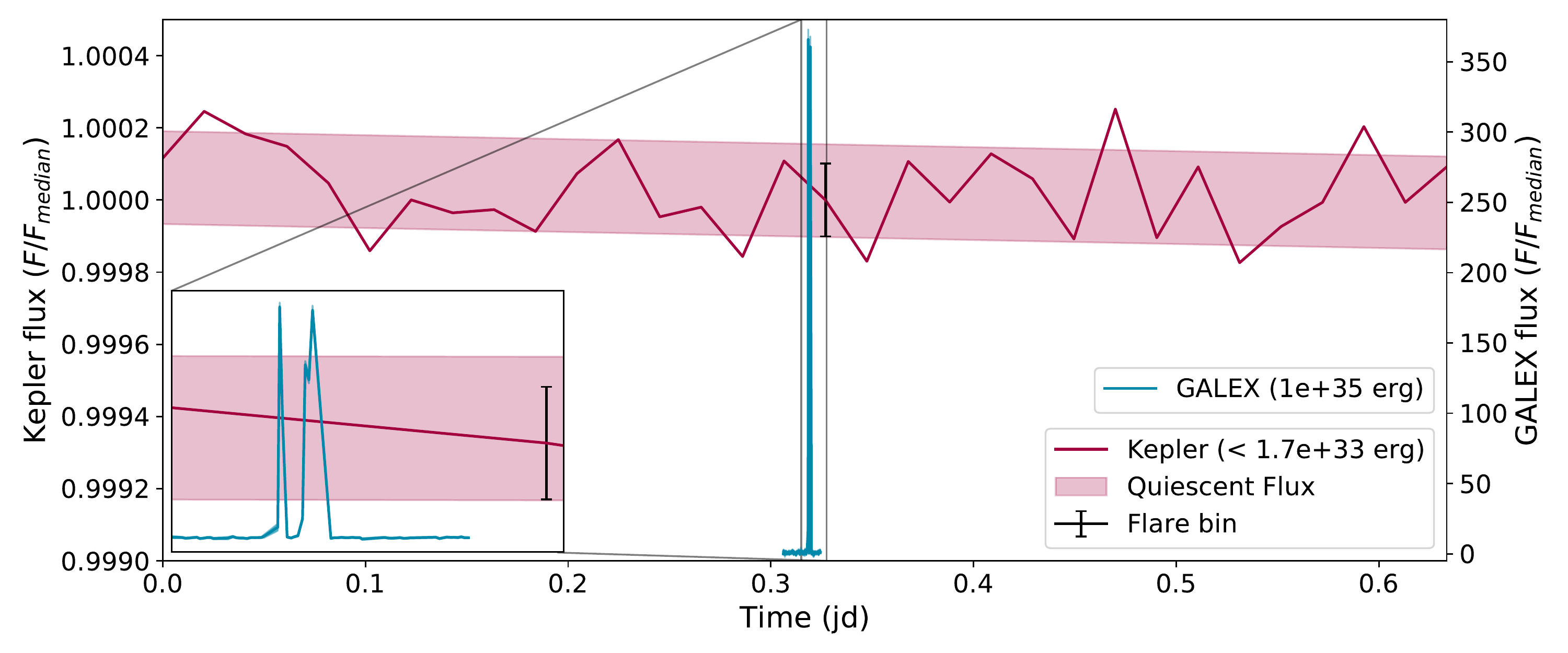}
    \includegraphics[width=.9\textwidth]{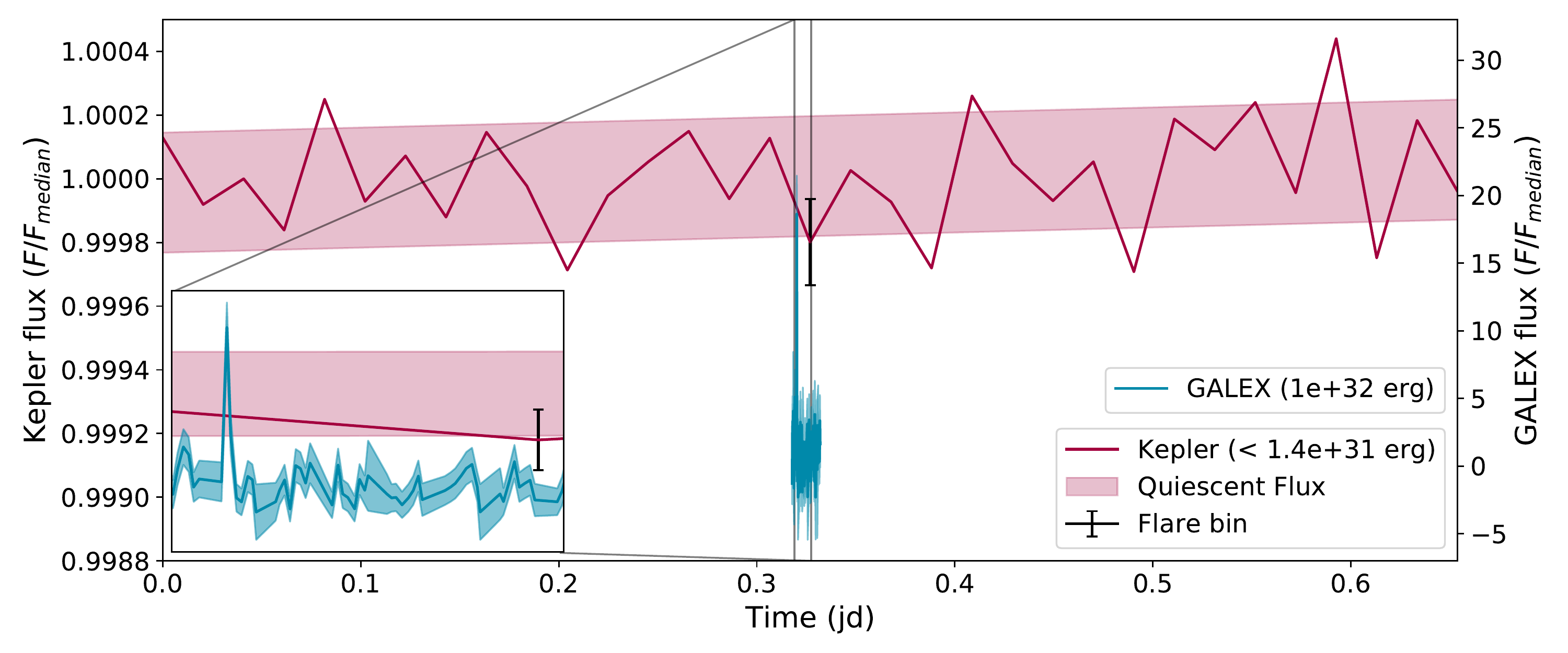}
    \caption{GALEX flares with associated Kepler non-detection. Each panel shows the GALEX and Kepler light curves at times of overlap; the inset to each panel zooms in on the GALEX light curve to examine the NUV flux variations on the much finer time scale on which it is recorded.
    The \textit{top} plot shows a $2.9\times 10^{36}$ erg NUV flare on KIC 9775887. The maximum Kepler band flare energy consistent with the Kepler light curve is $4.1\times10^{34}$ erg, two orders of magnitude less than we observe in the NUV. This is one of the largest NUV flares we observed.
    The \textit{middle} plot shows a $1\times 10^{35}$ erg NUV flare on KIC 9775956 (the second largest GALEX flare shown in fig \ref{fig:galex_kepler_contour}). The maximum Kepler band flare energy consistent with the Kepler light curve is $1.7\times10^{33}$ erg. This a medium energetic flare within our NUV flare sample.
    The \textit{bottom} plot shows a $1.0\times 10^{32}$ erg NUV flare on KIC 11668732. The maximum Kepler band flare energy consistent with the Kepler light curve is $1.4\times10^{31}$ erg. This is one of the smallest NUV flares we observed. 
    \label{fig:galexflarenokepler}
    }
\end{figure}

The Kepler long cadence light curves offered many more instances of data simultaneous to GALEX flares, however the large difference in cadence between the two data sets makes the analysis somewhat challenging. As expected (see discussion in the previous section) we were not able to detect any of our GALEX flares in the Kepler data, however what was not anticipated was not seeing any evidence of elevated flux in Kepler during any of the GALEX flare events.  Figure \ref{fig:galexflarenokepler} illustrates the situation well; it shows large (\textit{top}), medium (\textit{middle}), and small (\textit{bottom}) NUV flares with associated Kepler data. The medium flare can also be seen in Figure \ref{fig:galex_kepler_contour}. Note that for even the highest energy GALEX flare we cannot definitively say that the associated Kepler flux is elevated over quiescent flux. While this does not allow us to fully estimate the associated Kepler-band flare energy, it still allows us to place limits on the flare flux that can be produced in the Kepler bandpass. Table~\ref{tbl:monstertab} lists the stellar properties as well as the flare properties for all 1557 instances of a previously identified NUV flare and our constraints on the maximum energy flare in the Kepler band.

\begin{figure}[!h]
 \centering
    \includegraphics[width=0.8\textwidth]{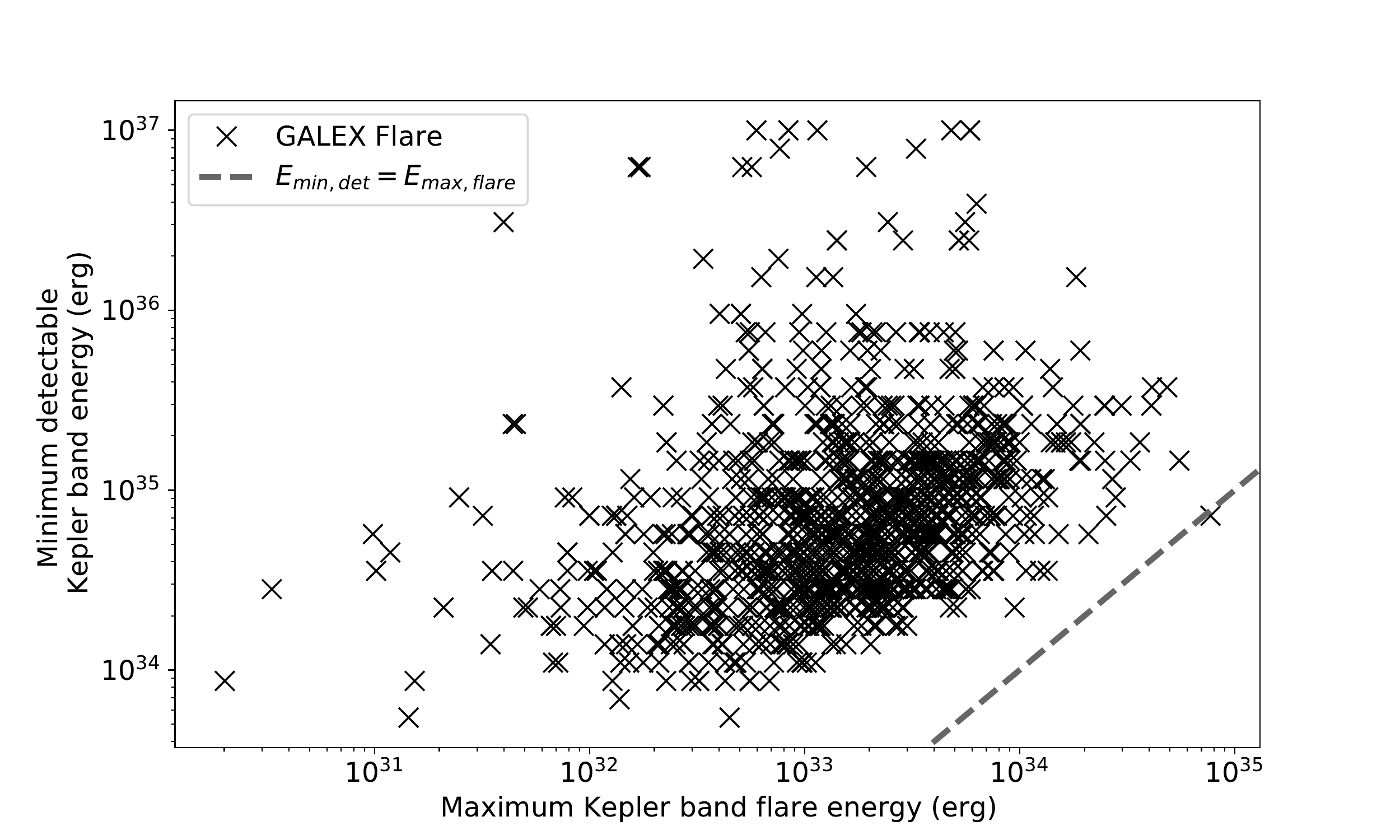}
    \caption{Plot comparing the  minimum detectable flare energy (as calculated through synthetic flare injection analysis, see Equation~\ref{eqn:recovery}) versus the maximum flare energy possible in the Kepler bandpass at the time of every simultaneous GALEX flare. The energies on both axes are the Kepler bandpass-specific energies, and the dotted line is the line of equality.
    The minimum detectable flare energy (ordinate of the plot) is a statistical measure that quantifies for each star the flare energy that would be detected 95\% of the time anywhere in the Kepler light curve using the assumptions of the flare detection algorithm described in \S\ref{sec:flarefinding}. 
    The maximum flare energy (abscissa of the plot) calculates excess emission over local quiescence in only the Kepler time bin corresponding to the time of the GALEX flare. 
    That there is almost two orders of magnitude difference in these two is largely due to incorporating the additional information about timing from the GALEX flares. 
    \label{fig:E_detectable_vs_E_max} }
\end{figure}

Given the short duration of the NUV flare in comparison to the Kepler data cadence, even if the flare is substantially longer in the Kepler band vs the GALEX band, we can be confident that the entirety of the flare flux enhancement occurs within a single time bin. Thus we do not need to make any assumptions about the flare shape in the Kepler band to place limits on its energy, given the observed flux. We calculated the maximum energy for the flare in the Kepler band given our observational data. To do this we first calculated the quiescent flux line around each GALEX flare data point. We then used that quiescent flux line with the single flare data point to determine the maximum flux above quiescence ($\Delta F$) for that data point given our uncertainties. Given this maximum flux enhancement for the flare as a whole, we used the distance to the individual star \citep{Bailer-Jones2018AJ....156...58B} to calculate the flare energy associated with that flux enhancement, making no assumptions about the shape of the flare. Figure~\ref{fig:E_detectable_vs_E_max} shows a comparison of the minimum detectable flare energy per star based on flare injection as discussed in \S\ref{sec:injection} with the maximum flare energy that could be undetected at the specific time of the GALEX flare dotted line is where the energies are equal.

\begin{table}[]
    \begin{tabular}{rrrrrrrr}
    KID$^{\dagger}$ & GID$^{\ddag}$ & Distance$^{1}$ & T$_{\rm eff}^{2}$ & Luminosity$^{3}$ & R$^{4}$ & E$_{\rm GALEX}$ & E$_{\rm max, Kepler}$ \\
      & & (pc) & (K) & (L$_{\odot}$) & (R$_{\odot}$) & (erg) & (erg) \\
      \hline
      11190862 & 3190134169448481571 & 5.1e+02 & 5994 & 9.10e-01 & 0.96 & 2.9e+33 & -- \\
      7102615 & 3190098985076398953 & 1.8e+03 & 3800 & 9.73e+02 & 71.08 & 6.1e+33 & 7.4e+35 \\
      7102399 & 3190098985076399645 & 9.4e+02 & 6040 & 5.58e-01 & 0.85 & 1.5e+33 & 2.2e+33 \\
      7183603 & 3190098985076400432 & 7.8e+02 & 4869 & 3.43e-01 & 0.82 & 5.2e+33 & 7.7e+32 \\
      6674968 & 3190098985076393704 & 5.7e+02 & 5891 & 5.84e-01 & 0.85 & 1.2e+33 & 1.0e+33 \\
      10212441 & 3190028616332221981 & 5.7e+02 & 5798 & 1.54e+00 & 1.21 & 6.0e+33 & 3.9e+32 \\
      9776769 & 3190028616332214740 & 8.8e+02 & 5342 & 4.75e-01 & 0.92 & 6.7e+33 & 5.3e+32 \\
      9959067 & 3190028616332216912 & 1.0e+03 & 6353 & 1.93e+00 & 1.31 & 3.5e+33 & -- \\
      9835972 & 3190028616332215404 & 4.6e+02 & 5847 & 8.32e-01 & 0.99 & 2.9e+33 & 2.0e+32 \\
      9776771 & 3190028616332214820 & 6.1e+02 & 5608 & 4.75e-01 & 0.83 & 1.1e+33 & 2.5e+32 \\
      \hline
      \multicolumn{8}{l}{$^{\dagger}$ Kepler Input Catalog ID number; $^{\ddag}$ GALEX Catalog ID number} \\
      \multicolumn{8}{l}{$^{1}$ Distance from \citet{Bailer-Jones2018AJ....156...58B};} \\
      \multicolumn{8}{l}{$^{2}$ Effective temperature from the Kepler Input Catalog (\url{https://archive.stsci.edu/kepler/kic.html});}\\
      \multicolumn{8}{l}{$^{3}$ Luminosity from \citet{Andrae2018AA...616A...8A}; $^{4}$ Radius from \citet{Andrae2018AA...616A...8A}}
    \end{tabular}
    \caption{Table listing characteristics of stars and flares with simultaneous UV and optical constraints. This table includes 1,557 GALEX flares with simultaneous optical constraints based on non-detections in the associated Kepler light curve. This table is published in its entirety in machine-readable format, while only a portion is shown here for guidance regarding its form and content.}
    \label{tbl:monstertab}
\end{table}
\FloatBarrier

\subsection{Kepler flares with overlapping GALEX data \label{sec:Kep_GALEX}}

As noted in Table~\ref{tbl:overlaps}, a small number of Kepler flares detected by the \citet{Yang_2019ApJS..241...29Y} survey had overlapping GALEX data. This group of seven objects is not a large enough sample to do a statistical analysis, but some simple statistics can reveal any commonalities with trends observed in other parts of the overlapping data. For each of the seven stars we divided the GALEX fluxes into fluxes that occurred while the star was undergoing a white-light flare according to \citet{Yang_2019ApJS..241...29Y} and fluxes that occurred during apparent quiescence. Table~\ref{tbl:Kepler_GALEX_overlap} lists the identities and stellar properties, as well as ratios of median and mean GALEX data during flares to GALEX data outside of Kepler flares. The ratio of medians is defined as
\begin{equation}
R_{\rm med} = \frac{median ({F_{\rm flare,i}})} {median ({F_{\rm nonflare,j}})}
\end{equation}
where $F_{\rm flare,i}$ are all the individual light curve flux measurements at the GALEX time binning of 10 seconds that occurred during flares in the Kepler band as noted by \citet{Yang_2019ApJS..241...29Y}, and $F_{\rm nonflare,j}$ are the individual GALEX light curve flux measurements which occurred outside any detected flares noted in the above study. Similarly, we define a ratio of means as
\begin{equation}
    R_{\rm mean} = \frac{\sum_{i} F_{\rm flare,i}/N_{i}} {\sum_{j} F_{\rm nonflare,j}/N_{j}}
\end{equation}
where $N_{i}$ are the number of GALEX fluxes occurring during Kepler flares, and $N_{j}$ are the number of GALEX fluxes occurring outside of Kepler-identified flares.

\begin{table}[]
    \begin{tabular}{lllcccccl}
       KIC &  T$_{\rm eff}^{\dagger}$ (K) & R$^{\ddag}$ (R$_\odot$)  & r$_{med}$ & r$_{mean}$  & $\Delta t_{\rm GALEX}/\Delta t_{\rm flare,Kepler}$ & E$_{\rm Kepler}^{*}$ (erg)& K-S & P  \\
       \hline
  3852116 & 4555 & 0.664 &  1.1 & 1.1 & 0.17 & 1.1e+34 & 0.19 & 0.35 \\
  7049035 & 4869 & 0.578 &  1.1 & 1.1 & 0.14 &1.4e+33& 0.12 & 0.24 \\
   11662738 & 4605 & 0.468  & 1.1 & 1.0 & 0.01 & 7.7e+33 & 0.19 & 0.37 \\
 7434110 & 3310 & 0.225  & 1.2 & 1.1 & 0.17 & 1.4e+32 & 0.31 & 0.0035 \\
  9268249 & 4864 & 2.245  & 1.1 & 1.1 & 0.11 & 5.6e+34 & 0.25 & 0.00042 \\
  11872364 & 5573 & 0.981 & 1.3 & 1.3 & 0.13 & 4.0e+34 & 0.45 & 4$\times$10$^{-14}$ \\
  6507888 & 3949 & 0.559  & 4.7 & 4.6 & 0.06 & 4.9e+33 & 0.92 & 2$\times$10$^{-30}$ \\
    \hline \\
    \multicolumn{9}{l}{$^{\dagger}$ T$_{\rm eff}$ from \citet{Gaia_2016AA...595A...1G,Gaia_2018AA...616A...1G} } \\
    \multicolumn{9}{l}{$^{\ddag}$ R from \citet{Berger_2018ApJ...866...99B}} \\
    \multicolumn{9}{l}{$^{*}$ Kepler flare properties from \citet{Yang_2019ApJS..241...29Y} }\\
    \end{tabular}
    \caption{Table of stellar properties with overlaps between Kepler LC flares and GALEX data}
    \label{tbl:Kepler_GALEX_overlap}
\end{table}

\begin{figure}[!h]
 \centering
    \includegraphics[width=\textwidth]{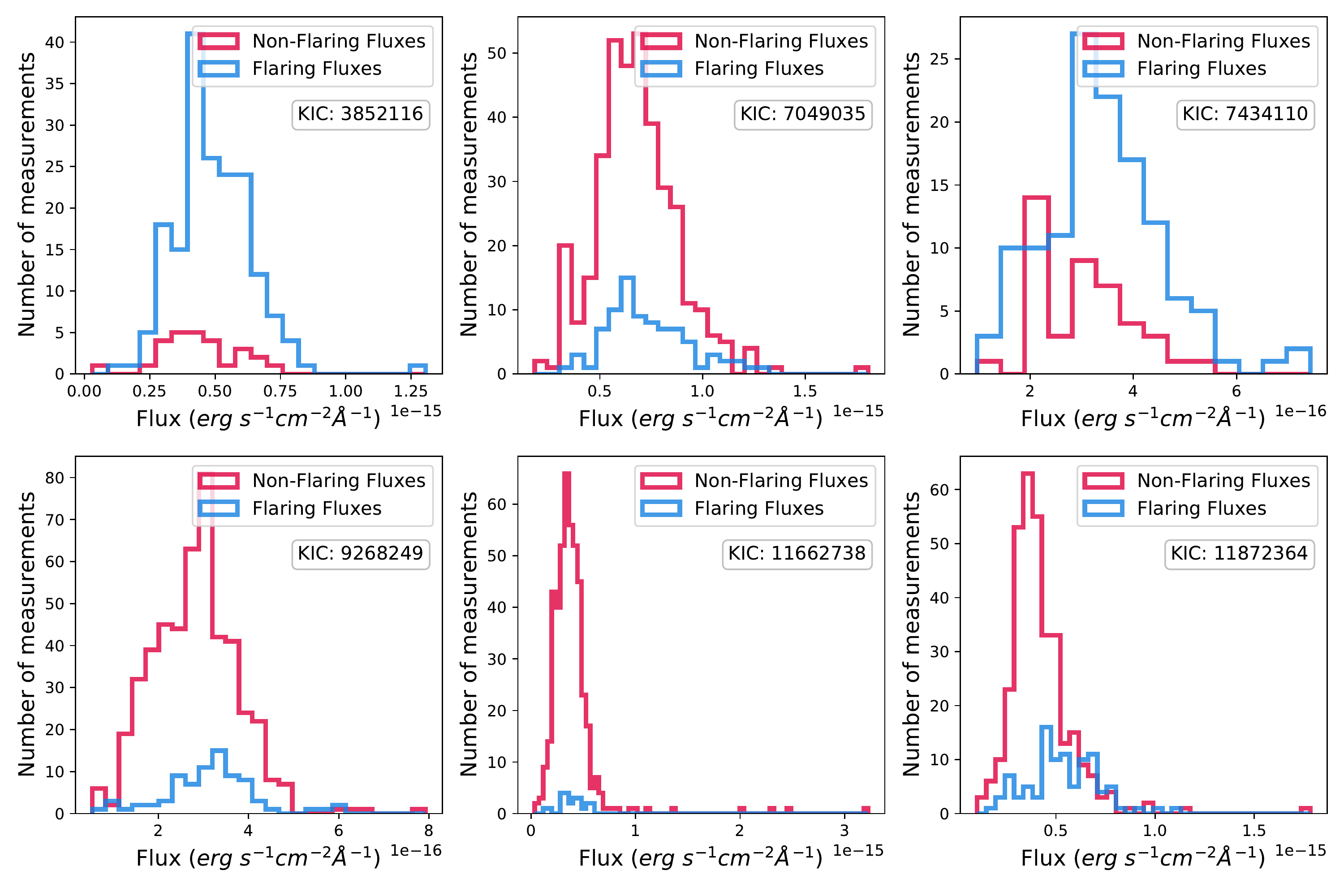}
    \caption{Histograms showing GALEX flux during and outside of Kepler flares for the six stars having overlapping data.
    \label{fig:galex_interval_histograms} }
\end{figure}

\begin{figure}[!h]
 \centering
    \includegraphics[width=\textwidth]{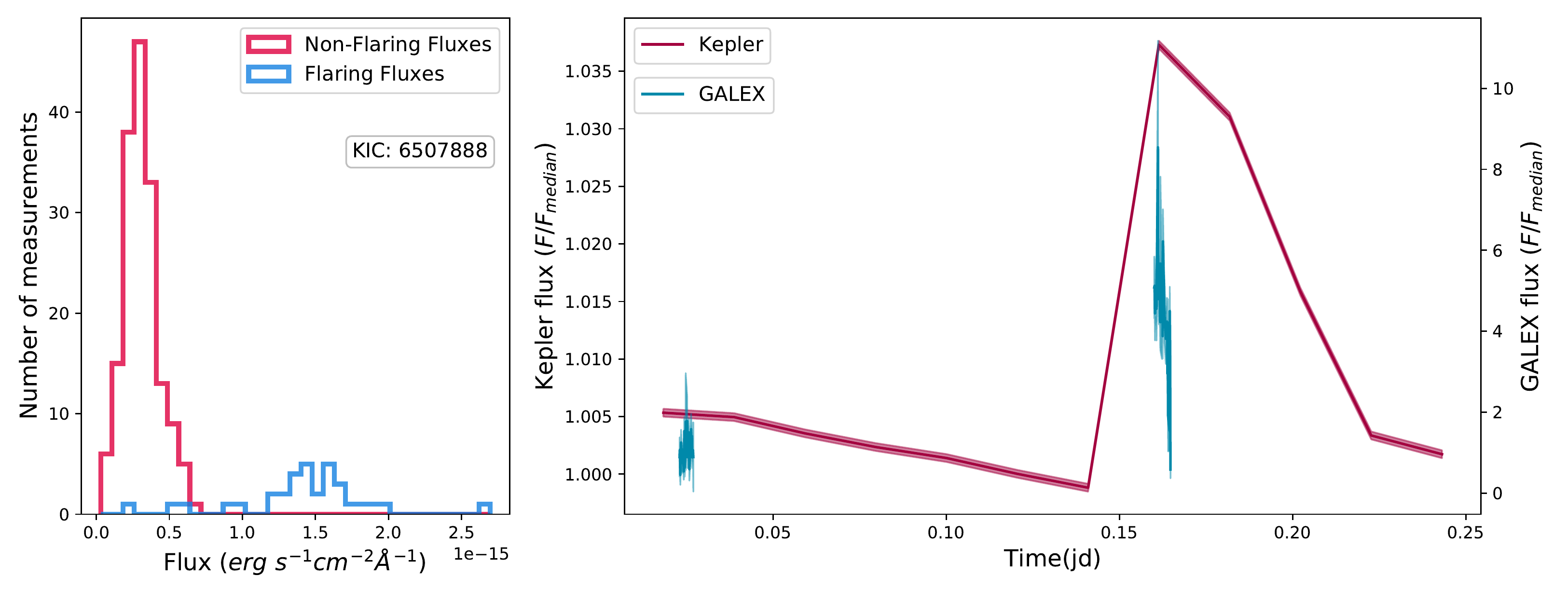}
    \caption{The figure displays the object where there is a large statistical difference in GALEX fluxes within vs. without Kepler flares. \textit{Left} panel indicates the histogram of data during and outside of flares, as in Fig.~\ref{fig:galex_interval_histograms}; \textit{right} panel displays the GALEX and Kepler data during the times of overlap. During the time of the Kepler white-light flare there may be an increase in NUV flux, but the short duration of the observations prevents a definitive comparison.
    \label{fig:galex_interval_outlier} }
\end{figure}

Figure~\ref{fig:galex_interval_histograms} displays histograms of six of the overlaps. We  performed Kolmogorov–Smirnov tests on each pair of (cumulative) distributions to determine whether evidence exists for the flux samples to be drawn from the same population or not. Only two of the Kolmogorov–Smirnov tests show a high likelihood that the flaring and non-flaring fluxes are drawn from different populations (KICs 6507888 and 11872364). One of the two shows only a slight elevation of the median and flux ratios, at 1.3 (KIC 11872364). The other result shows a clearly offset distribution between flaring and nonflaring fluxes. The histogram and underlying data for this object are displayed in Figure~\ref{fig:galex_interval_outlier}. The flare energies in the Kepler bandpass listed in  Table~\ref{tbl:Kepler_GALEX_overlap} span a large range but there is no systematic behavior in relating that to statistically significant NUV flux enhancements. The difference in time scale between the GALEX and Kepler data hampers this particular investigation, as the Kepler flares are so long compared to the GALEX data coverage that there is only overlap for a very small part of the flare.  The ratio of GALEX coverage during the Kepler flare to the flare duration is listed in Table~\ref{tbl:Kepler_GALEX_overlap}. The dissimilarity between the durations of Kepler and GALEX coverage translate into an inability to perform the reverse calculation of what was investigated in \S 3.3, e.g. limit the GALEX flare energy or flare flux increase for the range of Kepler observed flare energy or flare flux increase. Because the time intervals for GALEX are so much shorter, it is fairly certain that the GALEX data covers part of the Kepler flare, but not the other way around. 

\FloatBarrier

\subsection{Non-simultaneous flares in Kepler and GALEX}

\begin{table}
    \begin{tabular}{llllllll}
      KIC   & T$_{\rm eff}$$^{\dagger}$ & R$^{\ddag}$ & P$_{\rm rot}$$^{1}$ & N$_{\rm fl,Kepler}$$^{2}$ & MinMax E$_{\rm Kepler}$$^{2}$ & N$_{\rm fl,GALEX}$$^{3}$ & MinMax E$_{\rm GALEX}$$^{3}$ \\
       & (K) & (R$_{\odot}$) & (d) & & erg & & erg \\
 \hline
 3852071 & 5658 & 0.884 & - & 5 & 6.4e+33, 1.7e+34 & 1 & 8.5e+34, 8.5e+34 \\
4758595 & 3572 & 0.397 & - & 325 & 7.9e+30, 1.0e+33 & 2 & 1.9e+31, 1.9e+31 \\
7185248 & 4859 & 0.757 & 19.114 & 10 & 3.1e+32, 3.3e+33 & 2 & 7.1e+32, 7.1e+32 \\
8076634 & 5459 & 0.779 & 6.008 & 26 & 1.5e+33, 1.9e+34 & 2 & 1.4e+33, 1.6e+33 \\
8415404 & 4838 & 0.718 & 14.013 & 30 & 1.6e+32, 1.8e+33 & 2 & 2.4e+32, 4.4e+32 \\
9775887 & 5213 & 0.644 & 1.418 & 27 & 1.3e+33, 1.2e+34 & 9 & 8.5e+34, 2.9e+36 \\
10082757 & 6014 & 1.054 & 12.27 & 19 & 2.5e+33, 1.1e+34 & 2 & 2.4e+34, 6.0e+34 \\
10134076 & 6792 & 1.416 & - & 4 & 6.8e+33, 1.2e+34 & 2 & 7.8e+33, 4.9e+34 \\
10280703 & 4652 & 0.567 & 6.122 & 31 & 2.0e+32, 2.6e+33 & 1 & 2.1e+33, 2.1e+33 \\
10525463 & 4940 & 0.861 & 16.613 & 2 & 1.1e+33, 3.0e+33 & 1 & 1.9e+33, 1.9e+33 \\
11021136 & 5674 & 0.823 & - & 5 & 3.9e+32, 2.4e+33 & 1 & 3.8e+33, 3.7e+33 \\
11662738 & 4605 & 0.468 & 11.212 & 60 & 1.5e+32, 1.2e+34 & 1 & 3.1e+34, 3.1e+34 \\
 \hline
 \multicolumn{8}{l}{$^{\dagger}$ T$_{\rm eff}$ from \citet{Gaia_2016AA...595A...1G,Gaia_2018AA...616A...1G} } \\
 \multicolumn{8}{l}{$^{\ddag}$ R from \citet{Berger_2018ApJ...866...99B}} \\
 \multicolumn{8}{l}{$^{1}$ P$_{\rm rot}$ from \citet{McQuillanEtAl2014} } \\
 \multicolumn{8}{l}{$^{2}$ Kepler flare properties from \citet{Yang_2019ApJS..241...29Y} } \\
 \multicolumn{8}{l}{$^{3}$ GALEX flare properties from \citet{Brasseur_2019ApJ...883...88B}, for objects with a single flare, error }\\
 \multicolumn{8}{l}{~~~bars determine the min and max values.}\\
    \end{tabular}
    \caption{Table of stellar properties for flaring stars with non-overlapping GALEX and Kepler flares}
    \label{tbl:no_overlap}
\end{table}

\begin{figure}[!h]
 \centering
    \includegraphics[width=0.95\textwidth]{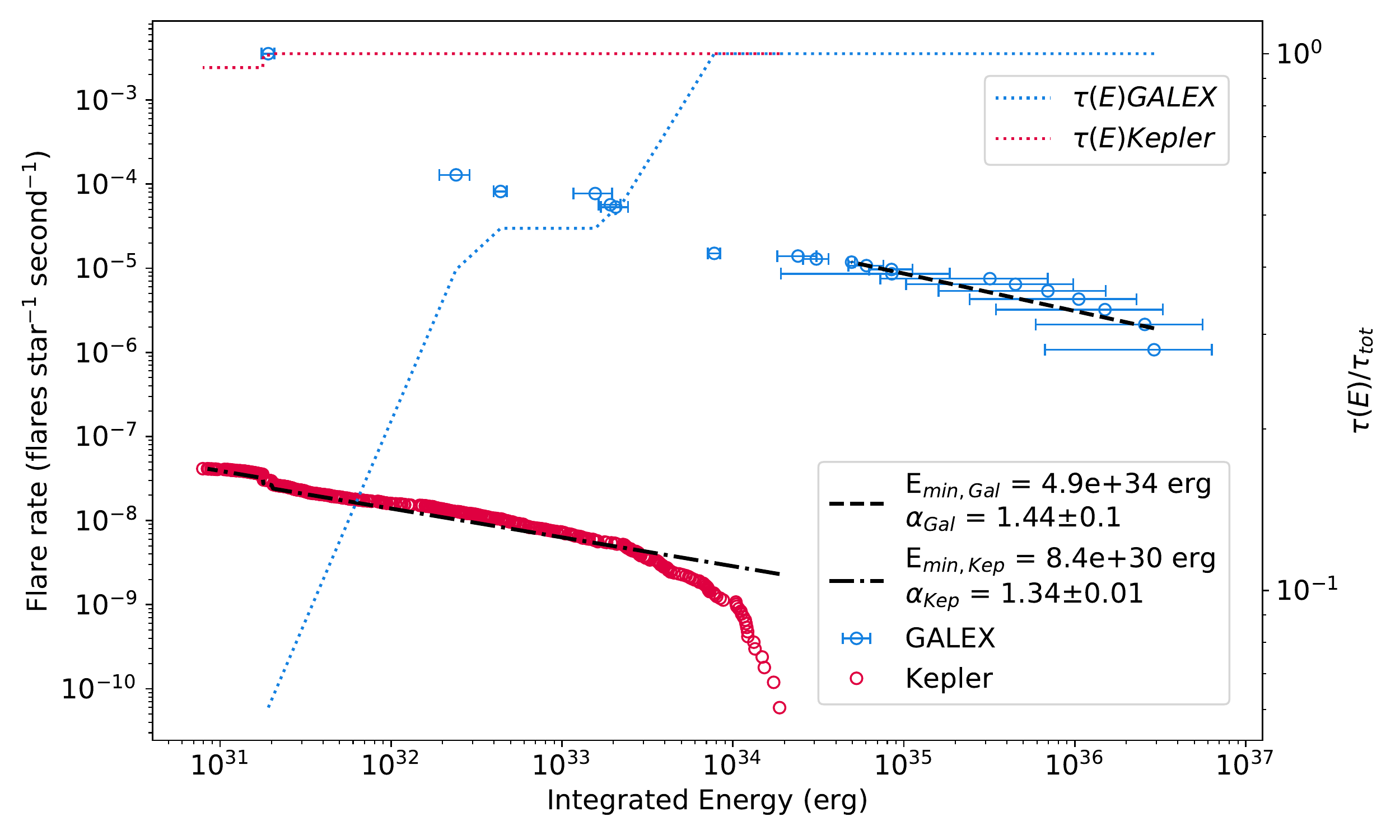}
    \caption{Flare rate comparison for the 12  stars which had GALEX NUV flare detections and Kepler long-cadence flare detections, although not simultaneously. The integrated energy on the abscissa is in the relevant bandpass.
    There is clearly a disconnect between the bandpass energies captured in the different flares, with NUV GALEX energies generally larger than the Kepler energies. Figure~\ref{fig:namekata} compares the energies and timescales of the Kepler and GALEX flares; while the flare energies span a common range, there is a large disconnect in the observed timescales. The axis on the right side shows the completeness curves for each data set; the percentage of observation time where flares of energy $E$ could have been detected based on the minimum detectable energy for that object. See \S\ref{sec:injection} for details on how the minimum detectable energy was calculated for Kepler light curves, and \citet{Brasseur_2019ApJ...883...88B} for the GALEX light curves.    \label{fig:flare_rate_comparision} }
\end{figure}

There are twelve stars which had flare detections in both the GALEX and Kepler bandpasses, albeit at differing times. Table~\ref{tbl:no_overlap} lists the identities of these stars, as well as some of their properties as gleaned from other catalogs. Flare rate distributions for both GALEX and Kepler flare populations followed that outlined in \citet{Brasseur_2019ApJ...883...88B}. Figure \ref{fig:flare_rate_comparision} shows the ensemble flare rate distribution for both GALEX NUV and Kepler bandpasses -- the integrated energy plotted here is the energy in the individual bandpasses, not the estimated bolometric energy (as that is one of the items under consideration in the present study).  In Figure~\ref{fig:flare_rate_comparision}, the dashed and dashed-dotted lines, respectively, delineate the range of flare energies above the minimum energy using a maximum likelihood approach to fit a power-law to the curve. The range of flare energies detected with GALEX covers a much larger range than observed with Kepler, and the ranges encompassed by the maximum likelihood fit to each dataset do not overlap. This offset may be a result of differing amounts of radiated energy appearing in these two bandpasses. The vertical offset in flare rate (flares per star per second) in Figure~\ref{fig:flare_rate_comparision} is partly explainable by the differing temporal resolutions of the two surveys (10 seconds vs 30 minutes) which cannot be entirely smoothed out by the conversion to flare rate. However this is also additional evidence that we are indeed seeing different phenomena in the two wavebands.  

From table \ref{tbl:no_overlap} we can see more than half of the Kepler flares come from a single object (KIC 4758595). If we remove this object from consideration, the resulting Kepler fits are $E_{min,Kep} = 2.5e+32$ erg and $\alpha_{Kep} = 1.54 \pm 0.04$. The change in minimum energy is due to KIC 4758595 having the majority of the lower energy flares in the sample. Removing this object from the GALEX sample changes the minimum energy much less ($E_{min, Gal} = 2.4e+34$ erg), while the $\alpha$ value changes similarly to the Kepler one, $\alpha_{Gal} = 1.39 \pm 0.1$. Thus overall we find that there is little bias introduced by the inclusion of a single object with such a higher flaring rate.

Although the two flare rate trends do not overlap either in energy range or flare rate range, the two $\alpha$ values which give the slopes of the power-law fit lines, do agree within the uncertainties. This result agrees with the findings in \citet[][and references therein]{Brasseur_2019ApJ...883...88B} when comparing flare occurrence rate distributions across different stellar types and different wavelength regions. These aggregate statistics  would seem to imply that a single energy fractionation could align these to a common energy scale. 

\section{Modeling} \label{sec:modeling}

As multiwavelength stellar flare studies become more prevalent, creating realistic models of flaring spectral energy distributions becomes increasingly necessary. While blackbody models have been able to replicate some aspects of flares, by definition they ignore spectral aspects like the Balmer jump, emission/absorption lines, and local deviations from the blackbody curve. By modeling the expected spectra from observations based on physical processes, we both gain an independent estimated energy fractionation to compare to observations and are able to gauge the differences between the processes in stellar and solar flares. The end result is a better physical understanding of the stellar habitat. In turn, observational studies inform models if the estimated fractionation is nonsensical.  Previous modeling of GALEX flare emission relied on CHIANTI thus assuming optically thin radiation \citep{Welsh2006}.  More realistic models are now possible with the time-dependent response of the low atmosphere, where optical depths in the GALEX and Kepler bandpasses achieve large values and are strongly wavelength dependent due to the increased densities in the flare chromosphere.   

We employ the RADYN 1D, plane-parallel radiative-hydrodynamic code \citep{Carlsson_1992ApJ...397L..59C,Carlsson_1995ApJ...440L..29C,Carlsson_1997ApJ...481..500C,Allred2015} to model the stellar atmospheric response due to the heating from  a nonthermal electron beam.   RADYN solves the time-dependent population densities and non-LTE radiative transfer simultaneously with the hydrodynamic equations \citep[for more details, see][]{Allred2015}. A  grid of models has been calculated with a large range of electron beam parameters.  The details of this grid and other applications are presented elsewhere \citep[][Kowalski et al. 2022b, in preparation]{Kowalski_2017ApJ...837..125K}.   Here, we focus on two flare models covering a large range of beam injection parameters in the grid.  The F13 model employs a maximum energy flux of $10^{13}$ erg cm$^{-2}$ s$^{-1}$ above a low-energy cutoff of 150 keV, and the 5F11 model has a maximum energy flux of $5 \times 10^{11}$ erg cm$^{-2}$ s$^{-1}$ above 25 keV. Both of these models use an injected electron power-law distribution; the F13 models employ $\delta=3$ and the 5F11 model employs a value of $\delta=4$. All of the beam parameters for the 5F11 were obtained from fits to RHESSI data of a large, well-observed X-class flare \citep{Kleint2016} that has been previously studied with RADYN models. The time-variation of the injected electron energy flux peaks at 1~s and ramps down until the calculation ends at 10~s.  This duration adequately encapsulates the observable response of the heating burst. Generally, the F13 is a much higher energy beam than those inferred in solar flares and is needed to reasonably replicate the optical and NUV response of M dwarf flares \citep{Kowalski_2015SoPh..290.3487K}.  

\begin{figure}[!h]
 \centering
    \includegraphics[width=.9\textwidth]{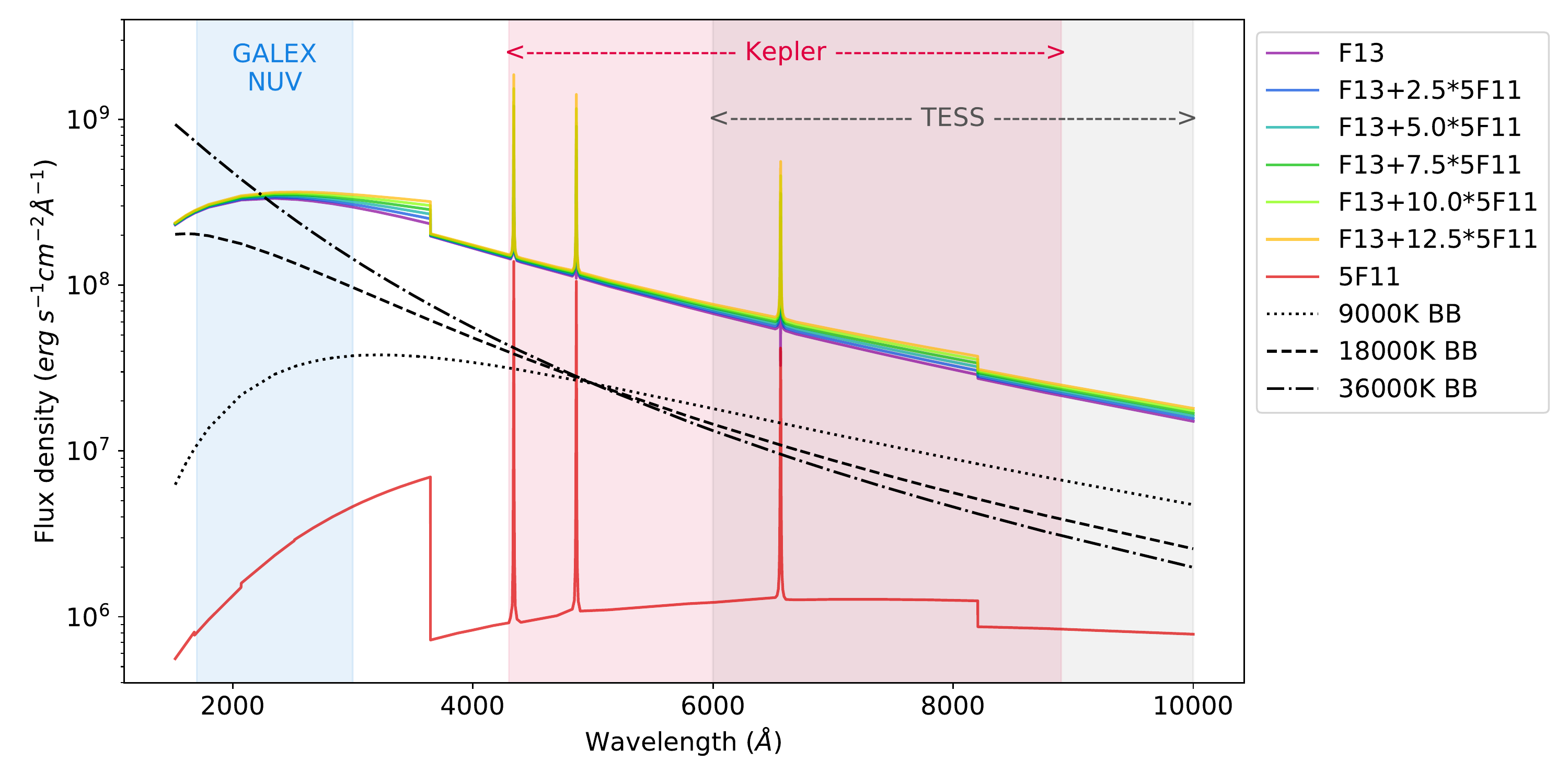}
    \caption{Average flux densities of several combinations of F13 and 5F11 models; each incorporates both continuum and the main H Balmer lines in the 1600 to 10000 \AA ~range and describes the response of a different distribution of accelerated particles in the stellar atmosphere. Section 4 explains the reasoning for the   different combinations of models. Blackbody models corresponding to a few characteristic temperatures (9000K BB, 18000K BB, 36000K BB) quoted in previous flare literature are shown for comparison. The Blackbody models all use a scale factor of 0.5 erg cm$^{-2}$ \AA$^{-1}$ s$^{-1}$ sr$^{-1}$.
    \label{fig:models} }
\end{figure}

The detailed radiative surface flux spectra averaged over the 10~s of heating are shown in Figure \ref{fig:models}.  We employ a combination of two models to represent conditions similar to what has been seen in solar flares \citep{Kowalski_2017ApJ...836...12K}. The two flare emission components may represent the spatially extended H$\alpha$-emitting ribbons and compact white-light kernels that are readily apparent in high spatial resolution images of solar flares \citep[\emph{e.g.}, as in Fig. 2 of][]{Kawate2016}. The F13 model spectrum shows enhanced blue continuum radiation with a color temperature $> 9000$ K, while the 5F11 exhibits a larger Balmer jump ratio. We combine the two models using a relative filling factor, $X_{\rm{5F11}} = 0 - 12.5$.  This superposition has the effect of increasing the Balmer jump ratio and Balmer line flux while maintaining a color temperature that is consistent with spectral observations of M dwarf flares in the blue optical wavelength regime. This superposition of two RHD models is analogous to the combination of a Planck function $T = 8000 - 13,000$ K and an F11 RHD model presented in \citet{Kowalski2010, Kowalski2012SoPh}.   The 5F11 model is included to better reproduce the observed flux in the hydrogen Balmer continuum and emission lines through a superposition with the F13 spectrum \citep{Kowalski_2022FrASS...934458K}.  Though this modeling approach is motivated  by M dwarf flare observations, similar simulations in a solar gravity predict, more or less, the same results within the resolution of the parameter space sampled by our grid \citep{Allred2006, KowalskiAllred2018}.  

Because of the time-averaged nature of the spectral flux density, the model ratios are both a flux ratio and an energy ratio (assuming the same time duration of the flare). From the combined model spectra, the pre-flare spectrum is subtracted and GALEX and \emph{Kepler} band fluxes are calculated by integrating the spectra with the filter response functions following the method of \citet{Sirianni_2005PASP..117.1049S}. The model GALEX/\emph{Kepler} flare-only energy ratios are shown as the gray bar in Figure \ref{fig:galex_vs_kepler}.  In calculating the flare energies, we have assumed a constant $X_{5F11}$ value over each flare, and we have multiplied by the FWHM of each effective area curve.  The ratios decrease from $1.167$ ($X_{\rm{5F11}}=0$) to $1.088$ ($X_{\rm{5F11}}=12.5$) corresponding to increasing the  area of the optically thick Balmer line and optically thin Balmer-continuum emitting ribbons.  Notably, the ratio calculated from the 5F11 flare-only spectrum is very low, $\approx 0.6$, which is surprising since this model exhibits a large Balmer jump in emission. 

\begin{table}[]
    \centering
    \begin{tabular}{c|c|c}
      Model   & GALEX/Kepler  ratio & GALEX/TESS  ratio\\
      \hline
    F13             & 1.17 & 2.09\\
    F13 + 2.5*5F11  & 1.15 & 2.04\\
    F13 + 5.0*5F11  & 1.13 & 1.10\\
    F13 + 7.5*5F11  & 1.12 & 1.96\\
    F13 + 10.0*5F11 & 1.10 & 1.92\\
    F13 + 12.5*5F11 & 1.09 & 1.89\\
    5F11            & 0.63 & 0.91\\
    9000K BB        & 0.39 & 0.62 \\
    18000K BB       & 2.49 & 5.06 \\
    36000K BB       & 5.57 & 12.83 \\
    \hline
    \end{tabular}
    \caption{Table of models and bandpass-specific flux (and energy) ratios  }
    \label{tbl:model_values}
\end{table}

Table~\ref{tbl:model_values} lists the values of the expected flux ratios in the GALEX to Kepler bandpass  for the different models presented in Figure~\ref{fig:models}. For this sequence of models, Table~\ref{tbl:model_values} also includes the expected flux ratios between GALEX and the TESS bandpasses, a question also explored by \citet{Jackman_2022MNRAS.tmp.2923J}. Clearly, additional complexity could be included in the flare model spectra in Figure \ref{fig:models}.   The blending of the high-order Paschen series will affect the Kepler bandpass to a small extent, while the Mg \textsc{ii} $h$ and $k$ resonance lines as well as a forest of Fe \textsc{ii} emission will affect the synthetic GALEX bandpass calculation \citep{Hawley1991, Hawley2007, Kowalski2019}.  Preliminary calculations indicate that in the 5F11 model, the emission line contribution is non-negligible, whereas the F13 is continuum-dominated.  These calculations are not included here, however, because there  are rather large discrepancies in the observed broadening (where spectra exist) and RHD models of non-hydrogenic emission lines in the NUV, such as Mg \textsc{ii}, in solar and stellar flares \citep{Hawley2007, Zhu2019}.  A detailed study of the effect of emission line forest with enhanced broadening in the GALEX bandpass will be presented in future work.  The RHD models in Figure \ref{fig:models} critically account for the time-evolution of the wavelength dependent continuum opacity spanning the two bandpasses, and thus improve significantly upon the widely used isothermal blackbody modeling approach.

\section{Discussion \label{sec:discussion} }

\subsection{Realistic stellar flare models \label{sec:disc2}} 

With the significant increase in monitoring the white-light flare activity of stars of a range of spectral type, it has become more standard to assume that a single Blackbody spectrum with a temperature of 9,000-10,000~K is appropriate to describe the flare emissions. Recent papers using both modeling and investigating systematics of observation interpretation reveal that this approach is overly simplified, even for M dwarfs where additional systematic effects due to stellar T$_{\rm eff}$ are minimized. For M dwarf stellar flare studies, the origin of this Blackbody model lies in early flare studies utilizing broad-band filter photometry across a range of wavelengths \citep{Hawley1991, Hawley_1995ApJ...453..464H} for which a Blackbody seemed like an appropriate fit. The advent of low resolution time-resolved blue-optical spectroscopy in M dwarf flare studies showed significant departures from this \citep{Kowalski_2012PhDT.......179K}. While there is a component of flare emission that can be approximately described by a blackbody of roughly 9,000K, the Balmer jump region exhibits significant emission  above this level, which both varies throughout the course of a flare as well as shows significant differences from flare to flare. More recently, \citet{Kowalski2019} combined NUV and blue-optical time-resolved spectroscopy for M dwarf flares and quantified that a Blackbody approximation underpredicts the total NUV continuum flare flux by about a factor of two and undepredicts the total NUV flare flux by about a factor of three. \citet{Kowalski_2013ApJS..207...15K} additionally noted the presence of an enhanced red continuum present in some M dwarf flares, which would contribute additional flux to flares studied in the Kepler and TESS bandpasses.

Despite the claim by \citet{Kretzschmar_2011A&A...530A..84K} finding a 9000 K blackbody in flares for the Sun and by implication, other solar-like stars, more recent studies have refuted this as a systematic effect of separating a flare blackbody from the stellar effective temperature blackbody \citep{Kleint2016, Kleint2020}. The proper analysis was first described using broadband filter observations of M dwarf flares in  \citet{Hawley_1995ApJ...453..464H} (see also Eq.\ 3 of \citet{Kowalski2016} and the appendices of \citet{Kowalski_2017ApJ...836...12K}), which does not assume an optically thin continuum source. For G-type stars, subtracting the $T_{\rm{q}} \approx 6000$ K quiescent spectrum from an optically thick flare spectrum represented by only a small temperature increase ($\Delta T \approx 500-1500$ K) in the photosphere is also consistent with a large ($T \gtrsim 8500$ K) color temperature.

\subsection{Short timescale flares \label{sec:shortt}}

While studies of the dynamics of stellar flares inherently involve the time domain, advances in flare studies in recent years have made great use of the flowering of interest in time domain studies more generally in the wider astrophysics community and the associated capabilities which have been developed. \citet{Kowalski2016} examined flare variations on one second timescales in the blue optical and optical wavelength ranges. The motivating paper for the current study, \citet{Brasseur_2019ApJ...883...88B}, examined a population of short-duration flares in the NUV using GALEX data at 10-second cadence. \citet{Webb2021} used 20 second cadenced optical imaging to study stellar flares within 500 pc across 12 fields of observations with the Dark Energy Camera (DECam), and found that the majority of flares occur on timescales less than 8 minutes, with a range of optical flare enhancements ranging from 0.1-1.8mag. \citet{Howard2021} reported on flare light curve profiles seen in the TESS bandpass at 20 second cadence, finding that roughly half of the large flares studied exhibited sub-structure in the rise phase of the flares. 

\begin{figure}[!h]
 \centering
    \includegraphics[width=\textwidth]{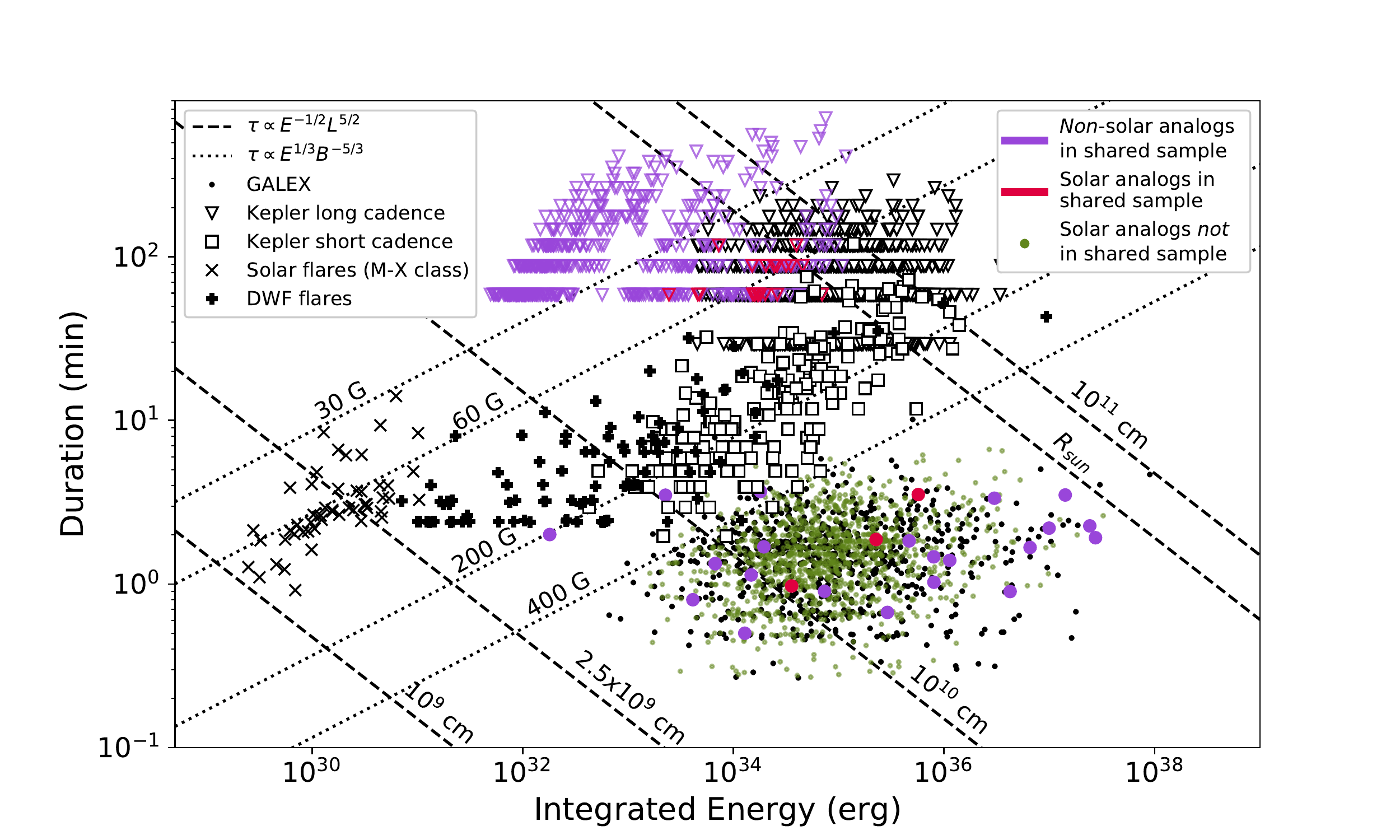}
    \caption{Plot of bolometric flare energy versus duration, originally from \citet{Namekata_2017ApJ...851...91N}, with data from \citet{Webb2021,Brasseur_2019ApJ...883...88B} and the present study added. Open squares and open triangles denote white light flares seen from solar-like stars in Kepler studies of short-cadence and long-cadence data, respectively, while ``x''es are solar white light flares. The pluses are \textit{g} band flares discovered in Deeper, Wider, Faster programme (DWF) data by \citet{Webb2021}. Black circles are the sample of GALEX NUV flares studied in \citet{Brasseur_2019ApJ...883...88B}, converted from NUV energy to bolometric in the manner described in that paper. Red/purple circles and triangles correspond to the subset of stars with non-simultaneous flares recorded in both GALEX NUV and Kepler bandpasses, respectively, with the colors representing solar (red) as described in \citet{Namekata_2017ApJ...851...91N} and non-solar (purple) analogs. Both red and purple circles overlap the energy and duration of the black circles, as expected since these flares were drawn from the \citet{Brasseur_2019ApJ...883...88B} study. Similarly the red triangles overlap the data reported in \citet{Namekata_2017ApJ...851...91N}, while the  purple triangles extend to shorter energies than the Kepler but retain the span of flare durations. This is because \citet{Namekata_2017ApJ...851...91N} only considered solar analogs. Dotted lines indicate expected trend of flare energy and duration if the emitting region length scale is held constant and the magnetic field varies; dashed lines indicate the expected trend for constant magnetic field and varying emitting region length scale. 
    \label{fig:namekata} }
\end{figure}

\citet{Namekata_2017ApJ...851...91N} presented a comparison of flare durations and energies for solar white light flares and white light flares seen on solar-like stars with Kepler. These data are reproduced in Figure~\ref{fig:namekata}, along with the GALEX NUV flare data examined in \citet{Brasseur_2019ApJ...883...88B}. The red and purple color coding in Figure~\ref{fig:namekata} indicates stars with flares observed in the Kepler long cadence data (open triangles) and GALEX NUV data (circles), although not observed simultaneously with both telescopes. Red color coding indicates stars that fit the descripton of ``solar-like'' in \citet{Namekata_2017ApJ...851...91N}, whereas purple color coding points out non-solar analogs in the shared sample. At the small energy end, the one smallest GALEX flare also has the lowest range of the Kepler band energy flares. There is a dispersion for more energetic GALEX flares. These also tend to be the ones that are the most non-solar, according to the criteria used by \citet{Namekata_2017ApJ...851...91N}. 

Figure~\ref{fig:namekata} encapsulates trends in flare duration and integrated energy across a sample of solar and stellar flares. While the UV and Kepler flares span roughly the same broad range of energy, there is a notable separation in flare duration. This significant offset arises partly from the difference in data collection, and is most noticeable between the GALEX flares and the long-cadence Kepler flares. The longest duration Kepler long-cadence flare lasts 260 minutes, and the shortest GALEX flare detected spans only 30 seconds. The short-cadence Kepler flares noted in \citet{Namekata_2017ApJ...851...91N} and displayed in Figure~\ref{fig:namekata} span the range of durations between these two extremes. We note that a similar plot of flare duration versus energy for the short duration optical flares in the \citet{Webb2021} sample (their Figure~8) shows a similar behavior as the Kepler short cadence flare characteristics. This evidence supports a continuum of flare durations between a few tens of seconds all the way up to several hours. The observed durations from \citet{Howard2021}, interpreted as the stop time of the flare minus the start time, appear to overlap those found from Kepler studies rather than revealing populations of short-duration flares similar to those seen with GALEX. The unusual HST/FUV flare recently noted by \citet{MacGregor2021} had an even shorter duration, at less than 10 seconds. 

The difference in behavior at NUV and optical wavelength regions suggests additional complexities to a common interpretation of the two wavelength regions as originating from the same physical processes. The forward slanted dotted lines in Figure~\ref{fig:namekata} indicate the expected trend of flare duration with energy for regions of constant magnetic field, whereas backward slanted dashed lines trace contours of constant length scale. These scaling laws come from equating the flare duration with the reconnection timescale,  associating the flare energy as a fraction of the magnetic energy released, and eliminating either the length scale or the magnetic field strength.  The first assumption, that the flare duration is related to the reconnection timescale, is likely to be in error, as the reconnection is taking place over a restricted region of space, which the length scales involved in the flaring structures can be much larger. As Kepler has a very red bandpass it is much more sensitive to the ``conundruum'' emission noted by \citet{Kowalski_2013ApJS..207...15K} than previous generations of pointed multi-filter optical flare monitoring. In the multithread flare scenario noted in \citet{Osten_2016ApJ...832..174O}, the lifetime of blue-optical flare emission was suggested to be shorter than the sustained red continuum emission. 

An additional ambiguity comes in the assumption of spatial configuration. Spatially unresolved stellar observations often assume a simple isolated flaring loop, whereas resolved solar flare observations reveal complexities such as arcades of loops and interconnecting active regions \citep{Warren2000}.
Each physical structure can have its own flare decay timescale related to length, density, temperature; furthermore, as described in \citet{Kowalski_2013ApJS..207...15K} different emission mechanisms will dominate in different wavelength regions and there is an added complexity of separate timescales for each process. Indeed, the flare template modeling of \citet{Davenport_2016ApJ...829...23D} uses two exponential light curve timescales during the flare decay phase, which may relate to the shorter timescale for the short wavelength continuum emission in the Kepler bandpass to decay versus  longer wavelength line emission which displays a longer decay timescale in spectral-temporal studies \citep{Kowalski2019}.

\subsection{Range of observed flare energy fractionation \label{sec:disc1}}

Most studies of stellar flares utilize a single bandpass for flare detection and characterization. An understanding of the spectral energy distribution of stellar flare radiation and bolometric flare energy estimates is needed for a holistic view of flare energetics and a better understanding of the impacts of stellar flares. Study of the energy fractionation, namely the fraction of the bolometric flare radiated energy that appears in different bandpasses, furthers progress in understanding flare processes.  Solar flare studies have the advantage of using bolometers to measure this integrated quantity directly \citep[e.g.,][]{Woods2006W}. Stellar work in this area relies on piecemeal approaches of examining flares in important parts of the electromagnetic spectrum for flare energy release. Early work in this area for M dwarf flares demonstrated the energetic importance of the blue-optical wavelength region during large M dwarf flares \citep{Hawley1991,Hawley_1995ApJ...453..464H}.  As discussed in \citet{Osten_Wolk_2015ApJ...809...79O}, the contribution of coronal flare radiation in the X-ray-Extreme Ultraviolet (XEUV)  also appears to have a significant minor contribution to the bolometric radiated flare energy. 

We make constraints on the range of observed flare energy fractionation via several different methods, elucidated in separate subsections herein: the constraints from simultaneous NUV and white light flare measurements in this paper; constraints from the literature for previously reported simultaneous NUV and white light flares; and nonsimultaneous NUV-white light flare constraints from this paper. 

\subsubsection{Simultaneous NUV-White Light Flare Constraints From this Paper \label{sec:sim_constraints}}

\begin{figure}[!h]
 \centering
    \includegraphics[width=\textwidth]{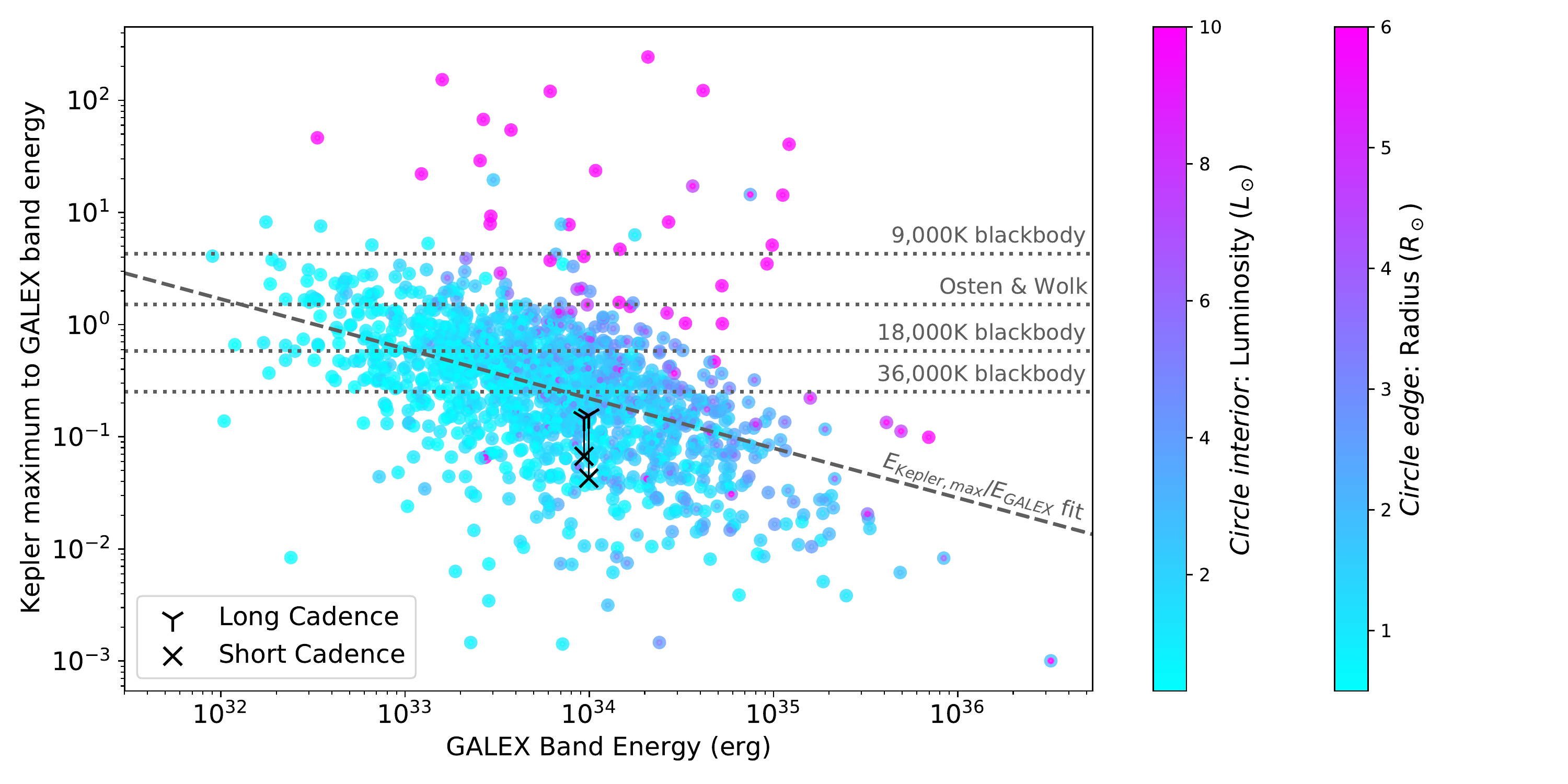}
    \caption{Energy fractionation for a sample of $\sim$1500 flares observed with GALEX and containing simultaneous Kepler LC data, against GALEX band flare energy. 
    Because of the mismatch in timescale sampled by GALEX vs. Kepler, we used the intensity in the Kepler data point which is strictly simultaneous with the GALEX flare to estimate the most energetic flare which would not be detected; see \S3.3 for details.
    Each point is color-coded by the stellar properties: the interior of each circle indicates the stellar luminosity, while the color of the circle's edge indicates the stellar radius. There is a clear decreasing trend in the energy fractionation as the NUV flare energy increases, shown with a dashed line. There are clear outliers above the main trend, which show up as largely evolved stars that escaped the initial cutoff criteria in \citet{Brasseur_2019ApJ...883...88B}. Horizontal dotted lines indicate the expected energy fractionation if a single temperature blackbody can explain the spectral energy distribution in both bandpasses. The two connected long/short cadence pairs show the two GALEX flares for which there was both long and short cadence simultaneous Kepler data (both on KIC 9592705). For both flares the short cadence maximum energy values are significantly lower than those calculated from the long cadence light curve. This indicates that for at least some of the light curves the Kepler energy upper limits are likely much higher than the true Kepler band energy.
    \label{fig:galeKepRatio_coloredOutliers} }
\end{figure}

Figure \ref{fig:galeKepRatio_coloredOutliers}  plots the (known) GALEX band flare energy vs. the ratio between  the maximum Kepler band energy for each flare and the GALEX flare energy for the sample of 1557 flares. The y-axis values being plotted are upper limits.   Three horizontal lines show the Kepler to GALEX band energy ratio for a 9,000 K, 18,000 K, and 36,000 K blackbody. Not only are none of these a good fit for the bulk of the flares, the data shows clear trend (dotted line) of decreasing ratio of Kepler to GALEX band energy ratios with increasing NUV flare energy. The points in Figure \ref{fig:galeKepRatio_coloredOutliers} are color coded in two ways, the interior is colored by stellar luminosity and the edge by stellar radius. This color coding reveals that most of the outliers with high Kepler to GALEX ratios are both very luminous and very large. These stars are most likely evolved stars that did not get removed from the sample earlier due to not having the V- or B-band magnitudes  used in prior color-magnitude cut-off. We therefor exclude these stars from our analysis and further plots.

Note that the maximum Kepler-band energy is not the same as the minimum detectable flare energy. This is because the minimum detectable energy is the measurement of the energy requirements for a flare to be detectable in any part of the Kepler light curve with no additional data, while the maximum Kepler-band energy for a flare non-detection is the maximum energy that is compatible with the Kepler observations during a known GALEX flare event. Figure \ref{fig:E_detectable_vs_E_max} shows the relation between the two quantities.

As described in \S\ref{sec:short_const}, we have GALEX-Kepler short cadence overlap for two flares. The measured and constrained energies in the GALEX and Kepler bandpasses are listed in Table~\ref{tbl:mwlit}. These are also plotted in Figure~\ref{fig:galeKepRatio_coloredOutliers}, where we can see that the short cadence light curves indicate a lower upper limit for the Kepler band flare energy than that indicated by the long cadence light curves. 

\subsubsection{Simultaneous NUV-White Light Flare Constraints from the Literature \label{sec:lit}}

There are only a handful of published examples of simultaneously obtained NUV and optical flares to use as a comparable sample for the phenomenon observed here. These data are summarized in Table~\ref{tbl:mwlit}. While the wavelength ranges do not overlap perfectly with each other or those used in our study, they are sufficient to allow for a investigation of order-of-magnitude inferences of energy fractionation, in particular to compare to the present datasets.

\citet{Hawley1991} reported on a large flare on the M dwarf star AD Leo observed in the FUV, NUV, and optical continuum bands including V and R; this is often termed the ``Great Flare" of AD~Leo. The integrated U-band energy was $\sim$6$\times$10$^{33}$ erg. There are caveats with interpreting the energy values from this flare, however, as the NUV data saturated during the most intense parts of the flare, and did not observe the flare in its entirety. From Table~6 of \citet{Hawley1991}, the NUV continuum, Mg~II and Ca~II line emission integrated over all time intervals and relevant wavelength regions in the flare indicate a lower limit to GALEX band-like energy of at least 9$\times$10$^{33}$ erg. Summing up the integrated energy in the V and R filters plus line emission from Hydrogen over all phases of the flare approximates roughly the Kepler bandpass, and produces an energy estimate similar in magnitude.

\begin{table}[]
    \centering
    \begin{tabular}{l|llc|cc|cc}
    \hline
    Flare & Star & Spectral  & Distance$^{*}$ &\multicolumn{2}{c}{NUV}   & \multicolumn{2}{c}{optical}   \\
    & & Type  & pc & bandpass (\AA\ ) & energy (erg) & bandpass (\AA\ ) & energy (erg) \\
    \hline
     HST-2$^{1}$ & GJ~1243 & M4V &11.97987$\pm$0.0052 &2444-2841  & 7$\times$10$^{30}$ & 3400-7400 & 2.3$\times$10$^{31}$ \\
     HST-1$^{1}$ &" & "& "&" & 1.3$\times$10$^{31}$ & " & 6.1$\times$10$^{31}$ \\
     \hline
      Great Flare$^{2}$ & AD~Leo&M3V&4.966$\pm$0.002 & 2000-3260,  & 9$\times$10$^{33}$$^{\dagger}$ & V+R+H lines & 1$\times$10$^{34}$ \\
         &  && & Ca~II+Mg~II &  & &  \\
    \hline
    GK-1$^{3}$ & KIC 9592705 &F7V& 414$\pm$4 & 1771-2831  &  $9.9\times 10^{33}$ & 4300-8900  & $<2.8\times 10^{32}$ \\
    GK-2$^{3}$ & "&"& " & "& $9.3\times 10^{33}$ & " & $<1.9\times 10^{32}$ \\
    \hline
    F2$^{4}$ & DG~CVn & M4V&18.3$\pm$0.1 & 1597-3480  & $\sim$2$\times$10$^{35}$ & V+R & 1.37$\times$10$^{35}$ \\
    \hline
     \multicolumn{8}{l}{$^{1}$ \citet{Kowalski2019}, $^{2}$ \citet{Hawley1991}, $^{3}$ this paper, $^{4}$ \citet{Osten_2016ApJ...832..174O}.}\\
    \multicolumn{8}{l}{$^{*}$ Distances taken from \citet{Gaia_2018AA...616A...1G}.} \\
    \multicolumn{8}{l}{$^{\dagger}$ As noted in \citet{Kowalski2019}, 
    the IUE LWP spectral observations of the Great Flare of AD Leo started }\\
    \multicolumn{8}{l}{at 1200 s after the flare began and are saturated; this is consequently a lower limit to the NUV energy.}\\
    \end{tabular}
    \caption{Table of flares with simultaneous constraints in the NUV
    and optical bandpasses, along with estimated flare energies in
    each bandpass. The data are sorted in increasing flare energy, and 
    include the two constraints from the current paper using
    GALEX and Kepler short-cadence data.}
    \label{tbl:mwlit}
\end{table}

\citet{Kowalski2019} recently reported on two NUV flares observed simultaneously with optical flares on the nearby M dwarf flare star GJ~1243 using NUV and blue-optical time-resolved spectroscopy. Energy calculations for these flares, dubbed HST-1 and HST-2, cover $\sim$12 minutes of the flare including the impulsive phase, and cover the same time intervals for NUV and optical spectral ranges. Aperture corrections for the 2444-2841 \AA\ spectral range are 14\% and originate from the 5s light curve in Figure 1 of \citet{Kowalski2019}. These two flares have the lowest energies of the sample, and overlap the energy range occupied by the largest solar flares.

A very large outburst on one of the two M dwarfs in the DG~CVn binary system, (one of two events discussed in \citet{Osten_2016ApJ...832..174O} and creatively named F2), had complete coverage in the V and R filters. Unlike the other three flares noted here, F2 did not happen as the result of a targeted flare campaign and so multi-wavelength coverage was subject to happenstance. A single measurement with the Swift UVW2 filter, which spans 1597-3480 \AA, occurred near the peak of the optical flare. Scaling the model from \citet{Osten_2016ApJ...832..174O}, which was constrained by both UVW2 and V+R points, to the V-band flux of F2, and additional integration in wavelength space, enabled an energy estimate for this flare. This is listed in Table~\ref{tbl:mwlit}. \citet{Osten_2016ApJ...832..174O} note that based on the V/R color temperature, the F2 event shows evidence for increased conundrum continuum emission, a phenomenon noted in \citet{Kowalski_2013ApJS..207...15K} and references therein to consist of enhanced red emission during the later phase of impulsive flares on M dwarfs. 

\begin{figure}[!h]
 \centering
    \includegraphics[width=.8\textwidth]{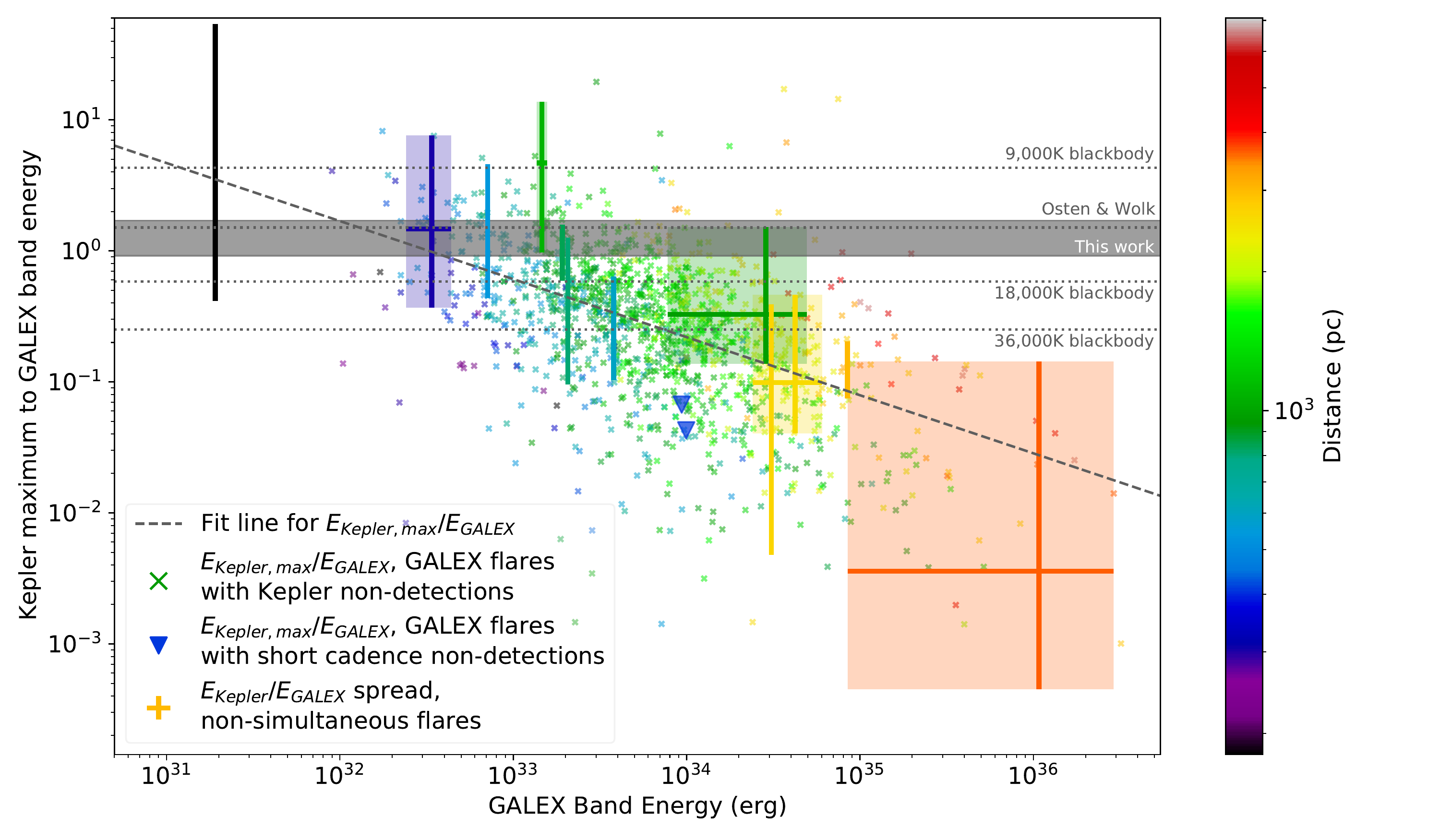}   
    \includegraphics[width=.8\textwidth]{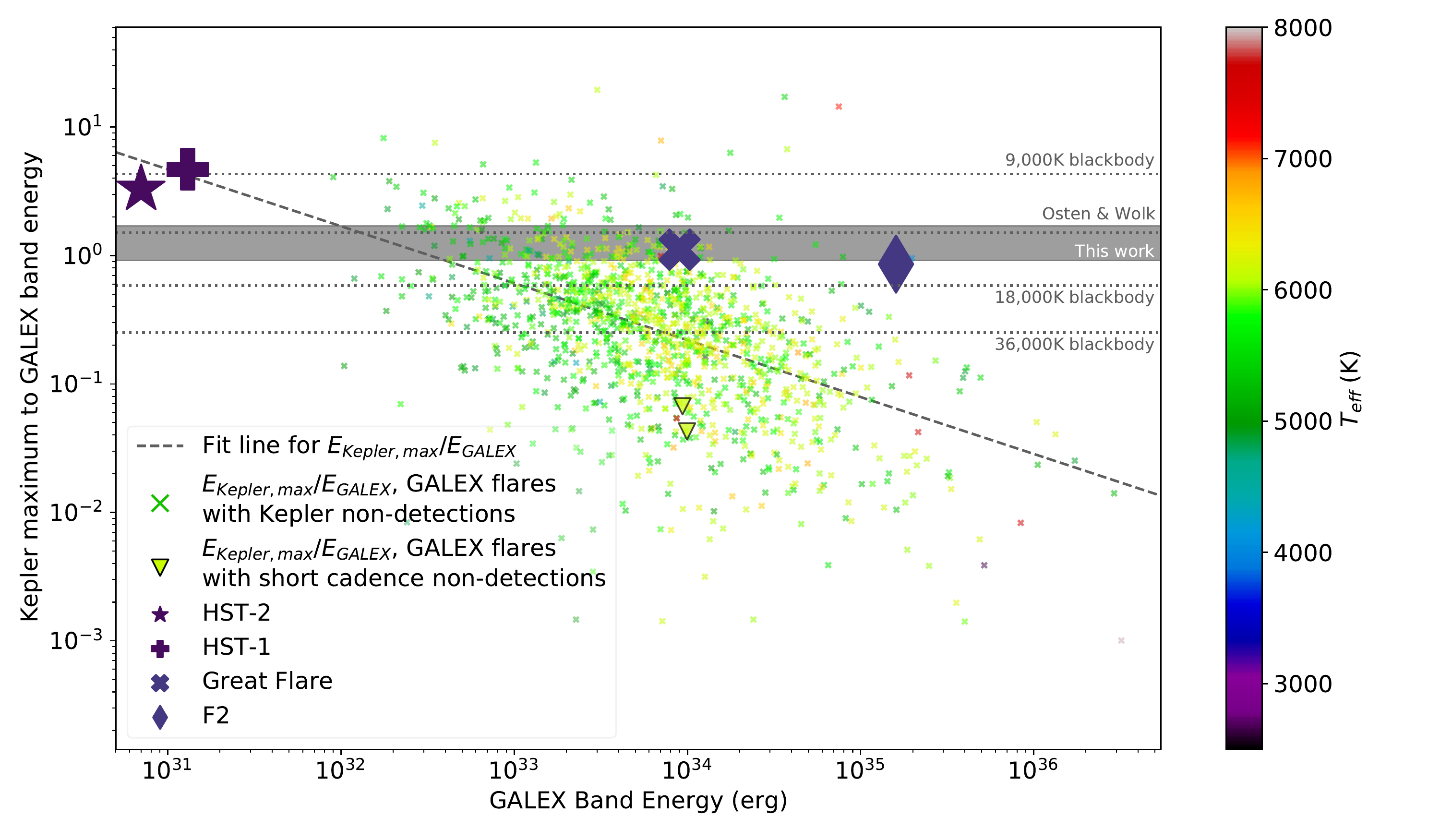}
    \caption{These plots synthesize the energy fractionation information from the present study relating flux in the the NUV bandpass with that from the optical bandpass. Each small x indicates the energy of a flare detected in the GALEX bandpass on the abscissa and the limit of the ratio between the maximum  undetected optical flare energy from long cadence Kepler data and this corresponding NUV flare energy, for simultaneously obtained measurements. The two GALEX flares with simultaneous constraints on optical flare energy from short cadence Kepler data are indicated with triangles. The dashed line is the fit to these upper limits as a function of NUV flare energy. Horizontal lines indicate various models for relating the expected energy fractionation between the two bandpasses, and are tabulated in Table~\ref{tbl:model_values} and discussed in \S4. 
    We would like to draw the reader's attention to the discrepancy between the two short cadence Kepler flares (triangles, discussed as GK-1 and GK-2 elsewhere in the paper) and the energy fractionation models, they are unequivocally an order of magnitude different than our RHD models.
    In the \textit{top} plot, also overplotted are the range of non-simultaneous flare energies measured in these bandpasses for stars which exhibited NUV and optical flares at different times. While the simultaneous and non-simultaneous data are not directly comparable, that we see the same trend in both indicates that there are bulk properties of the flares on a per star basis as a function of stellar type in addition to the individual flare to flare variation. Data in the \textit{top} plot are also color-coded by stellar distance. There is a clear bias in detecting the most energetic flares on the most distant objects. 
    The \textit{bottom} plot shows literature values of simultaneous data from NUV and optical flare campaigns. Table~\ref{tbl:mwlit} contains the details of these flares.  Data in the \textit{bottom} plot are also color-coded by effective temperature. The flares from the literature occur on stars which are all significantly cooler than the present sample and fit the model flare fractionation values better. This supports the idea that there are differences in physical mechanisms between flares on M dwarfs and earlier type stars.}
    \label{fig:galex_vs_kepler} 
\end{figure}

\subsubsection{Non-simultaneous NUV-White Light Flare Constraints from this Paper \label{sec:nonsim}}

The colored boxes in Figure~\ref{fig:galex_vs_kepler} \textit{top} mark the spread of  NUV and optical flare energies for non-simultaneous flare detections, with the plus marking the mean ratio (see table \ref{tbl:no_overlap} for a list of these objects). The trend of energy ratios for the non-simultaneous flares follows that of the data which have simultaneous constraints. The abscissa for both the simultaneous and non-simultaneous datasets is the measurement of the NUV GALEX energy. The ordinate in the case of the simultaneous datasets is an upper limit on the energy fractionation ratio. For the non-simultaneous datasets, however, the ordinate shows the spread of  measured Kepler flare energies, not a limit. The fact that the measurements overlap the upper limits is suggestive that this energy fractionation does vary over several orders of magnitude. For the ensemble of non-simultaneous datasets, this likely represents trends in the bulk properties of the flares and the different stellar types. \citet{Brasseur_2019ApJ...883...88B} noted a systematic influence of stellar distance on maximum flare energy observed, so this is another factor to consider. As noted above, the lowest energy GALEX flare observed non-simultaneously with Kepler flares occurs on a relatively nearby M dwarf, the only one in our study. From Table~\ref{tbl:no_overlap} the stellar properties of effective temperature and radius of these objects seem to be fairly representative of the initial selection criteria used to define the sample of  stars probed in the main Kepler mission. Rotation periods have been reported for 8 out of 12 of the objects \citep{McQuillanEtAl2014}, and all fall below 19 days indicative of enhanced magnetic activity. Visual inspection of Kepler light curves in the MAST archive for the others confirm an additional three with evidence for large-scale photometric modulations with periods less than 20 days, which supports the identification of NUV and white light flaring in these objects. 

\subsubsection{Synthesis of Data and Models, Inferences on Energy Fractionation}

Figure~\ref{fig:galex_vs_kepler} presents the compilation of results from simultaneous constraints on individual flare energy ratios described in \S\ref{sec:nondetections}, along with simultaneous constraints from the literature in \S\ref{sec:lit} and non-simultanious constraints in \S\ref{sec:nonsim}.  The results of the present study provide the largest compilation of constraints on flare energy fractionation across the widest range of stellar parameters and flare energy ranges. We find a dependence of the maximum optical/UV energy ratio distribution  on flare energy, which is a marked departure from the standard assumption that energy fractionation is a constant. Moreover, the range of this flare fractionation spans several orders of magnitude, becoming more pronounced for the more energetic flares in our sample.This appears to be at odds with the results from models and literature constraints from nearby M dwarf stars. This dependence is evident in both the maximum optical/NUV energy ratio from simultaneous constraints and the ratio of non-simultaneous stellar flare bulk properties for a sample of 12 stars with flares detected in the NUV and optical at different times. While the difference in data collection times prevents a more detailed examination, in 5 out of 7 Kepler flares with NUV coverage there is not a statistically significant elevation in NUV flux, and the other two display either a 30\% increase or a factor of 4 in NUV flux compared to quiescence. 

The four horizontal dotted lines in Figure~\ref{fig:galex_vs_kepler} indicate several flare energy fractionation estimates from models.  Three of these models assume a simple Blackbody curve with varying temperature to quantify the ratio of GALEX NUV band flare radiated energy to the flare radiated energy in the Kepler band. The commonly used 9,000K flare Blackbody temperature lies at the top, and does not overlap the bulk of the data. Other higher temperature Blackbody curves are shown, reflecting a factor of 2 and a factor 4 increase over the commonly used 9,000K model; the latter corresponds to a fairly high Blackbody temperature inferred from a flare observed in the FUV bandpass \citep{Froning_2019ApJ...871L..26F}.  \citet{Howard2020} also discusses higher flare Blackbody temperatures, finding a relationship between Blackbody temperature and flare energy. However, even a temperature of 36,000 K is not high enough to explain many of the optical/UV ratios we find, suggesting that the assumption of a single temperature Blackbody to describe the flare energy fractionation is too simplistic.

The dotted line marked ``Osten \& Wolk'' indicates the energy fractionation estimate from \citet{Osten_Wolk_2015ApJ...809...79O}. This estimate includes the quantification of both line and continuum radiation to the flare energy budget in the UV and optical, along with estimates of the contribution of coronal radiation to the total flare energy budget. This ratio is lower than that assuming only a 9,000K Blackbody contribution to the flare spectral energy distribution, but does not span the range of GALEX NUV to Kepler band energy ratios encompassed by our data. The black vertical line on the left side of the plot describes the several Kepler band flares and one GALEX NUV flare observed non-simultaneously on an M dwarf. Its flare energy fractionation range agrees much better with both the 9,000K blackbody value and \citet{Osten_Wolk_2015ApJ...809...79O}'s value. The flares from the M dwarf AD~Leo published by  \citet{Hawley_1995ApJ...453..464H} and used in the \citet{Osten_Wolk_2015ApJ...809...79O} derivation had U-band flare energies of a few times 10$^{32}$ erg, which is in the  range of the GALEX NUV flare energies where the trend line of the fit to the data crosses the \citet{Osten_Wolk_2015ApJ...809...79O} calculation, and suggests concordance with these fractionation values for M dwarfs. Most flare studies, particularly multi-wavelength flare studies, have historically been done on M dwarfs due to their frequent level of flaring. The values from the literature, plotted in Figure~\ref{fig:galex_vs_kepler} \textit{bottom}, show more agreement with the model values at a range of energies\footnote{There is an apparent discrepancy between the results shown here for HST-1 and HST-2 compared to the 9000 K blackbody model, and what is stated in \citet{Kowalski2019}, namely that these data indicate a factor of 2 disagreement between the continuum flux and what is predicted by a blackbody; this originates from the wavelengths at which the blackbody is normalized.}. The consistency of the results on M dwarfs suggests that the energy fractionation is related in part to stellar properties in addition to flare properties. More research will be needed  to tease apart these different effects using larger sample sizes of flares with simultaneous NUV and white light constraints spanning a range of stellar parameters.

The gray solid horizontal bar on Figure~\ref{fig:galex_vs_kepler} is the energy fractionation resulting from the models developed in the present study (see \S \ref{sec:modeling}). These models are more physically realistic and represent state of the art understanding of the influence of particle acceleration and plasma heating in the lower atmosphere during stellar flares. The literature values described in \S\ref{sec:lit} are more or less consistent with the models across a span of NUV flare energies. The model predictions are roughly consistent with the lower energy part of our flare energy trend, but still fail to account for the dependence of the flare ratio on energy. This underscores the importance of investigating more complex models that seek to more accurately represent the physical forces at play. 

There are a few factors to consider in seeking to understand the origin of the systematic dependence of flare fractionation on flare and stellar properties. The stars with literature values for simultaneous constraints (Table~\ref{tbl:mwlit}) are all M dwarfs exhibiting enhanced values of magnetic activity in general, and all less than 20 pc distant. They correspond to a generally young stellar population as evidenced by their frequent and extreme flaring events. We explored whether the significantly further distance of our sample stars could be imprinting a systematic effect on the determination of flare energies. The well-known wavelength dependence of interstellar reddening and extinction \citep{Cardelli1989} causes shorter wavelengths of light to be preferentially  absorbed and scattered, respectively, relative to optical wavelengths. As noted in \citet{Brasseur_2019ApJ...883...88B} and \S\ref{sec:energy}, standard values of reddening and extinction have been applied to both the NUV and optical data. While there are variations in the amount of interstellar reddening throughout our Galaxy, the typical values used for diffuse interstellar dust clouds  usually suffice. \citet{Brown2011} noted that the importance of this effect for stars in the Kepler field is fairly small. It would also work in the opposite sense to the the trend found in the present study, where we see an absence of white light emission during NUV flares. 

We explored the impact of flare detection statistics as a possible factor in this discrepancy.  \citet{Brasseur_2019ApJ...883...88B} noted that about 1810 out of 34,276, or 5\%, of their input sample contained some evidence for NUV variations after several iterations with different automated flare detection algorithms, which reduced further to 1145 stars   or 3\% after manual inspection of the light curves. The \citet{Yang_2019ApJS..241...29Y} Kepler flare catalog found 3420 flaring stars out of more than 200,000 stars searched, for an incidence of a little less than 2\%. These detection statistics are roughly comparable to each other, suggesting that the rate of short duration flaring in the NUV may only be elevated by about 30\% compared with the rate for longer cadence optical data. There are reasons to expect these numbers to be slightly different. The NUV data originate primarily from  lower-level magnetic activity-related variations and flaring emission, and this wavelength range also includes photospheric contributions. Optical wavelengths are dominated by emission from the photosphere, with flaring contributions added on top of that. Recently, \citet{Webb2021} reported on a 500 pc distance-limited sample of stellar flaring using 20-second cadence imaging at g-band (with 74 minutes on each field) as part of the Deeper Wider Faster flare search. This search is the most analogous in terms of cadence and time coverage to the NUV study which forms the basis of the present project. \citet{Webb2021} found 80 flaring stars out of 19,914 searched, for an incidence of 0.4\%, and noted the large preponderance of M dwarfs among their flaring stellar sample. \citet{Webb2021}'s study demonstrates that fast optical flares do occur. The flaring percentage noted in the NUV is higher than this by about an order of magnitude. The parameters of each study lead to different systematic effects: the input sample for the \citet{Brasseur_2019ApJ...883...88B} study is the Kepler Input Catalog, designed to favor solar-like stars over the more numerous low-mass M dwarfs. It spans a much wider range of distances. As noted above, we expect that the different contributions to the spectral energy distribution in each bandpass will affect flare detection statistics.

Examination of the maximum flare energy that could be undetected at the time of each GALEX NUV flare, as discussed in \S\ref{sec:lc_con}, is about two orders of magnitude lower than the minimum detectable flare energy derived after an analysis of injection and recovery of synthetic flares in the light curves in \S\ref{sec:injection}.  The artificially injected flares in the Kepler data were agnostic to flare physics, the process consisted of scaling a presumed flare profile in the Kepler band and integrating. The injection and recovery algorithms also considered statistics over the entire light curve and selected a threshhold such that the false discovery rate would be 10\%. As described in \S\ref{sec:injection}, for the ensemble of stars and flares considered, below about 10$^{34}$-10$^{35}$ erg most of the flare detections are false positives in the Kepler data. However we note that the current situation is the opposite: there is no optical flare to accompany an NUV flare. Given the general concordance of NUV flare rate with other flare studies of the same field, we do not find evidence for an enhanced  false positive NUV flare detection to affect our findings. The added constraint of matching the timing of the NUV flare enables a more sensitive constraint on the upper limit to the energy of a flare that could be occurring alongside the NUV enhancement, but undetectable within the particular flux level of the Kepler light curve at that time.

While there has been a resurgence of interest in the last few years in understanding how stars' UV and high energy environments affect planetary systems (especially during flares), most of the recent observations have concentrated on the FUV instead of the more energetically dominant NUV region. Still, there are some results which corroborate the extreme disconnects seen here:  \citet{MacGregor2021} reported on an FUV flare from the nearby M dwarf flare star Proxima with an unprecedented factor of more than 14,000 intensity increase above quiescence, at the same time as a mm flare with a factor of more than 1000 intensity increase above quiescence. Both were very short duration, lasting less than 10 seconds in total.  Simultaneous TESS observations did reveal a moderate flare with only a factor of $\sim$9 increase in intensity above the non-flaring value.  Examination of the rate of optical flares suggests a fairly common behavior ($>$75\% of white light flares having larger energies than that particular flare) while the extreme properties of the FUV flare mark it as a rare occurrence. This result confirms a potential disconnect between optical flare behavior and those at other wavelengths. While we cannot quantify the exact energy ratio in comparable terms relative to the flares discussed in the present paper, the TESS band energy was not significantly different from the FUV flare energy (10$^{31.2}$ erg vs 10$^{30.3}$ erg), and may even be in line with extrapolation of our trend to even lower flare energies. \citet{Loyd2020} noted the occurrence of two FUV flares on a relatively quiet M star planet host, GJ~887, over a 2.8 hr HST observation. Later observations spanning 28 days with TESS did not report any flares. One FUV flare was observed in its entirety with a bolometric energy of about 10$^{30}$ erg\footnote{The research note states that this is a bolometric flare energy but it is unclear what definition of bolometric is used; it may refer to the contribution of line and continuum flare radiation from the lower atmosphere quantified in \citet{Hawley2003}, however we note that an alternate definition of bolometric flare energy in \citet{Osten_Wolk_2015ApJ...809...79O} includes the contribution of coronal radiation.}. The two datasets are not simultaneous, unfortunately; it is not clear what the sensitivity of the TESS data is as regards TESS band flare energies that could be occurring but not detectable. \citet{Jackman2021} discussed the disconnect in UV and white light flares between non-simultaneous GALEX and TESS flare data for low mass stars. They found that the predicted NUV flare activity based on TESS data was roughly a factor of five below that observed, although no simultaneous constraints were at hand.

An examination of both Figures~\ref{fig:namekata} and \ref{fig:galex_vs_kepler} reveals that the flare duration is a key distinguishing factor between the NUV flares and Kepler flares. However, it is not the reason for the significant difference between the detected NUV flare energy and the upper limit on flare enhancement in the Kepler light curve at the same time. As noted in \S\ref{sec:shortt}, the range of flare durations span a wide gamut and from our modelling in \S\ref{sec:modeling} there is no reason to expect that short duration NUV flares would not have an optical counterpart. The two examples of NUV flares occurring during short-cadence Kepler light curves provide even more stringent limits, and it becomes difficult to understand how the flare spectral energy distributions discussed in \S\ref{sec:modeling} are compatible with the data in Figure~\ref{fig:galex_vs_kepler}.

\subsection{Implications of the results for understanding stellar radiative impact on exoplanets}

The explosion in precision optical photometry of stars in the last decade or so has resulted in a vast increase in understanding of optical stellar flare rates and the diversity of properties. Because flares are known to be multiwavelength in nature, it is natural to take the optical flare measurements and attempt to constrain the UV flux incident on exoplanets as a result of stellar flaring. Recent examples include \citet{Howard2020}, who used 2 min cadence Everyscope observations of optical flares in the g band to estimate the UV flux by extrapolating from the optical observations. \citet{Feinstein2020} used optical flare rates from TESS to estimate the NUV irradiation of planets. 

The results previously published by \citet{Kowalski2019} indicated that extrapolating a 9000 K Blackbody that is fit to the observed blue-optical continuum flux underestimates the NUV continuum flare flux by a factor of two and is a poor approximation of the full $\lambda = 3400-7500$ \AA\ continuum distribution. This was for two examples of blue-optical flares accompanied by NUV flare enhancements. 

The model calculations in \S\ref{sec:modeling} provide flux ratios between the GALEX-Kepler and GALEX-TESS bandpasses under a standard flare scenario relating the action of accelerated particles and subsequent heating in the photosphere and chromosphere. These should be used for more accurate estimation of UV flux enhancements under different conditions in the flare. Note that these flux ratios are for well-behaved flares, of which we see few examples in our data. Additional work is needed to account for the previously noted red optical enhancement in some flares \citep{Kowalski_2013ApJS..207...15K}. 

Our results suggest that  calculations based on white light flare measurements can only place a lower limit on the actual amount of UV flux incident on a planet due to the assumption about the relationship between the UV and optical flares. The present data harmonize with recent results suggesting a more complex relationship between the different stellar atmospheric layers than previously appreciated, as well as the possibility of new physics at play in stellar flares. This discrepancy should give pause to any further attempts to calculate the UV irradiation of planets using only the white light flaring rate, as this will dramatically underestimate the actual value. 

We find only a fairly small range of NUV flux enhancements during white light flares, but this is limited largely by the very low number of coincidences between white light flares and NUV data coverage. Of 7 Kepler band flares, only two show a statistically significant increase in NUV flux, with one increasing by 30\% and the other by a factor of 4. 

\section{Conclusion} \label{sec:conclusion}

We present the largest compilation to date of simultaneous constraints on flares occurring at NUV and optical wavelengths, over a wide variety of cool stars largely unbiased towards high flaring rates. This data reveals a large spread in the ratio of flare energy emitted in the NUV to optical bandpasses over 4 orders of magnitude in NUV flare energy.  The flare energy fractionation at low flare energies (which encompasses both solar flare energy ranges and the low energy end of M dwarf flares) and for a range of flare energies observed on M dwarfs does appear to be consistent with model expectations for the fraction of a flare's radiative energy to appear in the NUV and Kepler bandpasses. The most extreme flares in our sample, which originate from  G and K stars, display a disconnect of at least three orders of magnitude between the NUV flare energies observed and the constraints on maximum white light flare energy which could remain undetected in the simultaneous data.

In two out of 7 white light flares (table \ref{tbl:Kepler_GALEX_overlap}, figures \ref{fig:galex_interval_histograms} and \ref{fig:galex_interval_outlier}) we are able to infer evidence for NUV flux increase at the 30\% up to factors of several level during white light flares. The very different durations of the GALEX and Kepler data prevent us from being able to say anything more quantitative relating the increase in NUV flux during white light flares. 

We find that the ensemble slope  of flare frequency distributions for stars which flared in both the NUV and white light are consistent with each other, even as the individual bandpass energies probed in each are disjoint. This might suggest that the distributions can be aligned using a common energy partition to overlap each other, however considering the span of flares in each star reveals that these non-simultaneous multi-wavelength flares exhibit the same trend of energy fractionation between NUV and white light with NUV flare energy as deduced from constraints using simultaneous observations. 

We present more realistic models to relate the energy emitted in the NUV and white light bandpasses; these are to be preferred over simplistic Blackbody profiles. Even with a range of scenarios for beam heating and filling factor, the models predict a small range of energy fractionation, and  do not cover the span of observational constraints retrieved in this study. This range in model results is higher than  the spread in energy fractionation predicted from assuming a Blackbody distribution over a factor of four in temperature from 9000 - 36,000 K. Putting all the pieces of our analysis together we see that the simple 9000K Blackbody model often used to describe stellar flares is not adequate to explain the spread in energy fractionation, and furthermore no single energy fractionation model independent of the flare properties (principally flare energy but also possibly stellar type and other factors) is sufficient. Additional work to understand the response of a flaring atmosphere to different amounts of flare energy is clearly needed. 

These results are consistent with the confusing picture emerging about the disconnect between UV and optical ranges in terms of flare energy partition seen in some M dwarf flare studies. Our sample, in contrast, skews toward solar-like stars and stars with a low flaring rate. The large number of flares with simultaneous constraints cements the conclusion that despite decades of ultraviolet observations of flaring stars, we are still revealing the depths of our ignorance in understanding the diversity of flare responses. This motivates additional studies to understand the relation between the two bandpasses and what the implications are for stellar flare models. 

The results of this paper imply that the properties of an observed optical flare cannot be used to predict the UV flare radiation contribution  if it was not observed at the same time. This is a setback in the statistical approach to understanding the impact of flare radiation from planet-hosting stars on exoplanets, and points to the need to probe this on a more individual basis, both for each star and for each flare. 

\acknowledgements
All of the {\it Kepler} and {\it GALEX} data presented in this paper were obtained from the Mikulski Archive for Space Telescopes (MAST): \dataset[10.17909/T9CC7G]{http://dx.doi.org/10.17909/T9CC7G}, \dataset[10.17909/T98304]{http://dx.doi.org/10.17909/T98304}. STScI is operated by the Association of Universities for Research in Astronomy, Inc., under NASA contract NAS5-26555. Support for MAST for non-HST data is provided by the NASA Office of Space Science via grant NNX09AF08G and by other grants and contracts.
This paper includes data collected by the Kepler mission. Funding for the Kepler mission is provided by the NASA Science Mission directorate.
This work has made use of data from the European Space Agency (ESA) mission {\it Gaia} (\url{https://www.cosmos.esa.int/gaia}), processed by the {\it Gaia} Data Processing and Analysis Consortium (DPAC, \url{https://www.cosmos.esa.int/web/gaia/dpac/consortium}). Funding for the DPAC has been provided by national institutions, in particular the institutions participating in the {\it Gaia} Multilateral Agreement.
This work made use of the gaia-kepler.fun crossmatch database created by Megan Bedell.\\
The authors would like to thank the referee for their thoughtful and thorough comments.

\facilities{ GALEX (NUV), Kepler}

\software{Astropy \citep{astropy:2013,astropy:2018}, gPhoton \citep{gphoton_ref}, LightKurve \citep{LightKurve}, Astroquery \citep{astroquery}, appaloosa \citep{Davenport_2016ApJ...829...23D}, dustmaps \citep{dustmaps}, NumPy \citep{numpy}, SciPy \citep{scipy}, Matplotlib \citep{matplotlib}}

\bibliography{Kepler_GALEX_Flares,Software}
\bibliographystyle{aasjournal}

\end{document}